\documentclass[american,english,prl, twocolumn]{revtex4-1}
\usepackage[T1]{fontenc}
\usepackage[latin9]{inputenc}
\usepackage{color}
\usepackage{array}
\usepackage{float}
\usepackage{bm}
\usepackage{amsmath}
\usepackage{amssymb}
\usepackage{graphicx}

\makeatletter
\providecommand{\tabularnewline}{\\}

\usepackage{amsmath}
\usepackage{subfigure}
\usepackage{graphicx, mathtools}
\usepackage[colorlinks=true,linkcolor=MidnightBlue,urlcolor=MidnightBlue,citecolor=MidnightBlue,anchorcolor=MidnightBlue]{hyperref}\usepackage[dvipsnames]{xcolor}
\newcommand{\myparallel}{{\mkern3mu\vphantom{\perp}\vrule depth 0pt\mkern2.5mu\vrule depth 0pt\mkern3mu}}

\makeatother

\usepackage{babel}
\begin{document}

\title{Universal hydrodynamic mechanisms for crystallization in active colloidal
suspensions}

\author{Rajesh Singh}
\email{rsingh@imsc.res.in}

\affiliation{The Institute of Mathematical Sciences-HBNI, CIT Campus, Chennai
600113, India}

\author{R. Adhikari}
\email{rjoy@imsc.res.in}

\affiliation{The Institute of Mathematical Sciences-HBNI, CIT Campus, Chennai
600113, India}
\begin{abstract}
The lack of detailed balance in active colloidal suspensions allows
dissipation to determine stationary states. Here we show that slow
viscous flow produced by polar or apolar active colloids near plane
walls mediates attractive hydrodynamic forces that drive crystallization.
Hydrodynamically mediated torques tend to destabilize the crystal
but stability can be regained through critical amounts of bottom-heaviness
or chiral activity. Numerical simulations show that crystallization
is not nucleational, as in equilibrium, but is preceded by a spinodal-like
instability. Harmonic excitations of the active crystal relax diffusively
but the normal modes are distinct from an equilibrium colloidal crystal.
The hydrodynamic mechanisms presented here are universal and rationalize
recent experiments on the crystallization of active colloids.\\\\DOI: \href{https://journals.aps.org/prl/abstract/10.1103/PhysRevLett.117.228002}{10.1103/PhysRevLett.117.228002}\\ 
\end{abstract}
\maketitle
In active colloidal suspensions \cite{palacci2013living,petroff2015fast},
energy is continuously dissipated into the ambient viscous fluid.
The balance between dissipation and fluctuation that prevails in equilibrium
colloidal suspensions \cite{einstein1905theory,kubo1966fluctuation}
is, therefore, absent. Nonequilibrium stationary states in active
suspensions, then, are determined by both dissipative and conservative
forces, quite unlike passive suspensions where detailed balance prevents
dissipative forces from determining phases of thermodynamic equilibrium.
In this context, it is of great interest to enquire how thermodynamic
phase transitions driven by changes in free energy are modified in
the presence of sustained dissipation. 

In two recent experiments disordered suspensions of active colloids
have been observed to spontaneously order into two-dimensional hexagonal
crystals when confined at a plane wall. Bottom-heavy synthetic active
colloids which catalyze hydrogen peroxide when optically illuminated
are used in the first experiment \cite{palacci2013living} while chiral
fast-swimming bacteria of the species \emph{Thiovulum majus }are used
in the second experiment \cite{petroff2015fast}. Given this remarkably
similar crystallization in two disparate active suspensions it is
natural to ask if the phenomenon is universal and to search for mechanisms,
necessarily involving dissipation, that drive it.

Our current understanding of phase separation in particulate active
systems is derived from the coarse-grained theory of motility-induced
phase separation (MIPS) where active particles are advected by a density-dependent
velocity \cite{tailleur2008statistical,cates2010arrested,cates2013active,cates2015}.
Microscopic models with kinematics consistent with MIPS also show
phase separation and crystallization of hard active disks have been
reported in two dimensions \cite{henkes2011active,fily2012athermal,bialke2012crystallization,redner2013structure}.
However, these models ignore exchange of the locally conserved momentum
of the ambient fluid with that of the active particles and are, thus,
best applied to systems where such exchanges can be ignored. Fluid
flow is an integral part of the physics in \cite{palacci2013living,petroff2015fast}
and a momentum-conserving theory, currently lacking, is essential
to identify the dissipative forces and torques that drive crystallization. 

In this Letter we present a microscopic theory of active crystallization
that connects directly to the experiments described above. Specifically,
we account for the \textit{three-dimensional} active flow in the fluid
and the effect of a plane wall on this flow. Representing activity
by slip in a thin boundary layer at the colloid surface \cite{ghose2014irreducible,singh2014many,singh2016traction}
we rigorously compute the long-ranged many-body hydrodynamic forces
and torques on the colloids. Thus we estimate Brownian forces and
torques to be smaller than their active counterparts by factors of
order $10^{2}$ (for synthetic colloids in \cite{palacci2013living})
to $10^{4}$ (for bacteria in \cite{petroff2015fast}) making them
largely irrelevant for active crystallization. We integrate the resulting
deterministic balance equations numerically to obtain dynamical trajectories.

\begin{figure}
\includegraphics[width=0.45\textwidth]{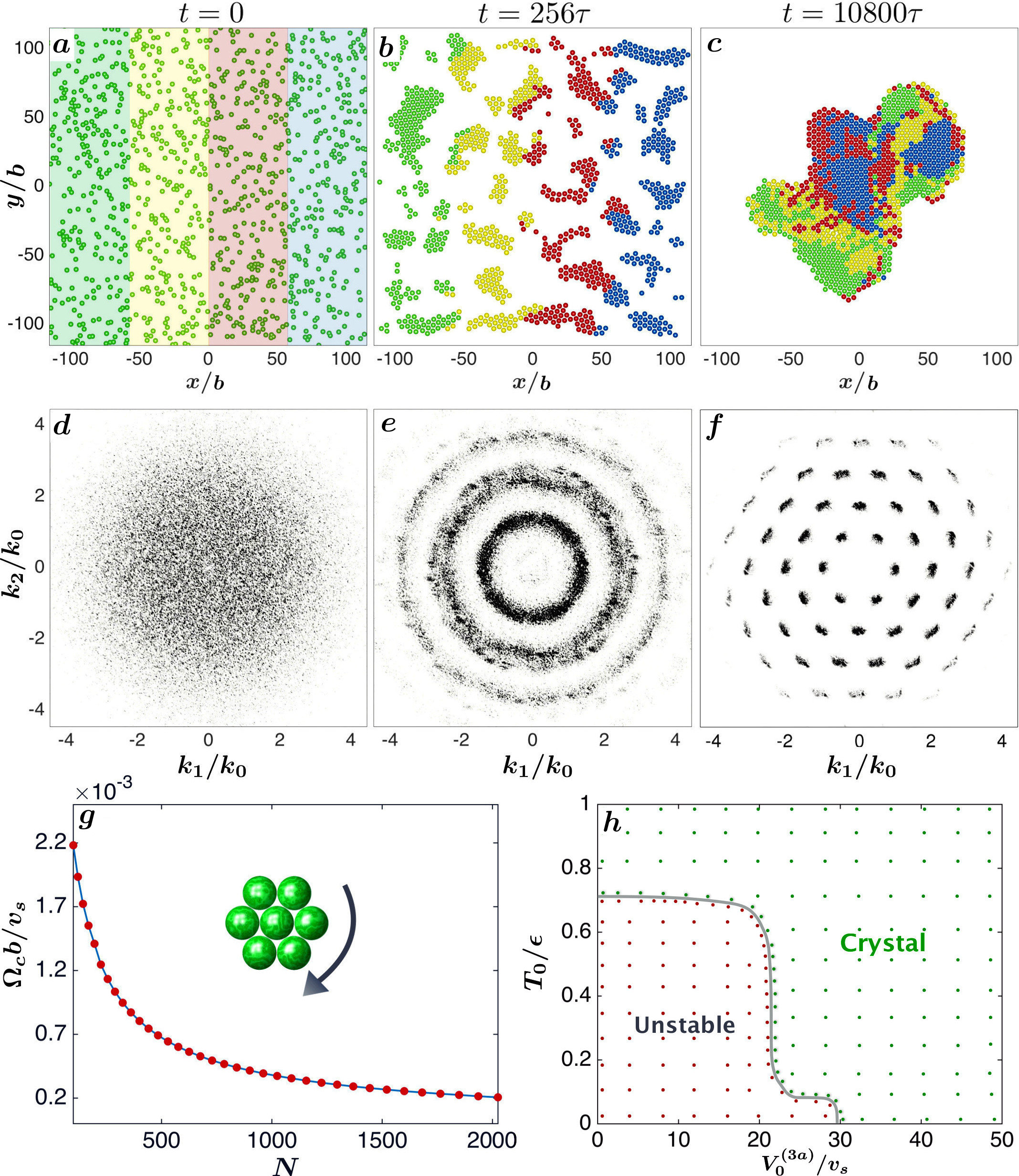}\caption{Panels (a)-(c) are instantaneous configurations during the crystallization
of $1024$ active colloids of radius $b$ at a plane wall. The colloids
are colored by their initial positions. Panels (d)-(f) show the structure
factor $S(\mathbf{k})$ at corresponding instants. Wavenumbers are
scaled by the modulus of the reciprocal lattice vector $k_{0}$ and
the contribution from $\mathbf{k}=0$ is discarded. Panel (g) shows
the variation of the angular velocity $\mathbf{\Omega}_{c}$ of a
crystallite with the number $N$ of colloids in it. A typical configuration
is shown in the inset. Panel (h) is the state diagram for orientational
stability in terms of the measure of chirality $V_{0}^{(3a)}$ and
bottom-heaviness $T_{0}$ (see text). Each dot represents one simulation.
Here $v_{s}$ is the self-propulsion speed of an isolated colloid,
$\tau=b/v_{s}$, and $\epsilon$ is the scale of the repulsive steric
potential.\label{fig:Dynamics-of-crystallization}}
\end{figure}

Our main numerical results are summarized in Fig. (\ref{fig:Dynamics-of-crystallization}).
Panels (a)-(c) show the spontaneous destabilization of the uniform
state by attractive active hydrodynamic forces, the formation of multiple
crystallites, and their coalescence into a single hexagonal crystal
at late times. Panels (d)-(f) show the structure factor at corresponding
times. The route to crystallization is not through activated processes
that produce critical nuclei, but through a spinodal-like instability
produced by the unbalanced long-ranged active attraction. The uniform
state is, therefore, always unstable and crystallization occurs for
all values of density, in contrast to the finite density necessary
for crystallization in MIPS models \cite{cates2015}. Active hydrodynamic
torques tend to destabilize the ordered state but stability is regained
when these are balanced by external torques (from bottom-heaviness
in \cite{palacci2013living}) or by chiral activity (from bacterial
spin in \cite{petroff2015fast}). Crystallites of chiral colloids
rotate at an angular velocity that is inversely proportional to the
number of colloids contained in them, as shown in panel (g). This
is in excellent agreement with the experiment \cite{petroff2015fast}.
The critical values of bottom-heaviness and chirality above which
orientational stability, and, hence, positional order, is ensured
is shown in panel (h). We now present our model and detail the derivation
of our results. 

\emph{Model:} We consider $N$ spherical active colloids of radius
$b$ near a plane wall with center-of-mass coordinates $\mathbf{R}_{i},$
orientation $\mathbf{p}_{i}$, linear velocity $\mathbf{V}_{i}$,
and angular velocity $\boldsymbol{\Omega}_{i}$, where $i=1\ldots N$.
Activity is imposed through a slip velocity $\mathbf{v}_{i}^{\mathcal{A}}$
which is a general vector field on the surface $S_{i}$ of the $i$-th
colloid satisfying $\int{\bm{\hat{\rho}}_{i}\cdot}\mathbf{v}_{i}^{\mathcal{A}}\,d\text{S}_{i}=0$
\cite{slipConstraint}, where $\boldsymbol{\rho}_{i}$ is the vector
from the center of the colloid to a point on its surface. The fluid
velocity $\mathbf{v}$ is subject to slip boundary conditions 
\begin{equation}
\mathbf{v}(\mathbf{R}_{i}+\boldsymbol{\rho}_{i})=\mathbf{V}_{i}+\bm{\Omega}_{i}\times\bm{\rho}_{i}+\mathbf{v}_{i}^{\mathcal{A}}(\bm{\rho}_{i}).\label{eq:slip-RBM-BC}
\end{equation}
on the colloid surfaces, to a no-slip boundary condition $\mathbf{v}=0$
at the plane wall located at $z=0$, and to a quiescent boundary condition
at large distances from the wall. The slip is conveniently parametrized
by an expansion $\mathbf{v}^{\mathcal{A}}(\mathbf{R}_{i}+\boldsymbol{\rho}_{i})=\sum_{l=1}^{\infty}\tfrac{1}{(l-1)!(2l-3)!!}\,\mathbf{V}_{i}^{(l)}\cdot\mathbf{Y}^{(l-1)}(\bm{\hat{\rho}}_{i})$
in irreducible tensorial spherical harmonics $\mathbf{Y}^{(l)}(\bm{\hat{\rho}})=(-1)^{l}\rho^{l+1}\bm{\nabla}{}^{(l)}\rho^{-1}$,
where $\bm{\nabla}^{(l)}=\bm{\nabla}_{\alpha_{1}}\dots\bm{\nabla}_{\alpha_{l}}$.
The expansion coefficients $\mathbf{V}_{i}^{(l)}$ are $l$-th rank
reducible Cartesian tensors with three irreducible parts of ranks
$l,$ $l-1,$ and $l-2$, corresponding to symmetric traceless, antisymmetric
and pure trace combinations of the reducible indices. We denote these
by $\mathbf{V}_{i}^{(ls)}$, $\mathbf{V}_{i}^{(la)}$ and $\mathbf{V}_{i}^{(lt)}$
respectively. The leading contributions from the slip,
\begin{eqnarray}
\mathbf{v}_{i}^{\mathcal{A}}(\bm{\rho}_{i}) & = & \underbrace{-\mathbf{V}_{i}^{\mathcal{A}}+\tfrac{1}{15}\mathbf{V}_{i}^{(3t)}\cdot\mathbf{Y}^{(2)}}_{\mathrm{achiral,\,polar}}-\underbrace{\tfrac{1}{9}\boldsymbol{\varepsilon}\cdot\mathbf{V}_{i}^{(3a)}\cdot\mathbf{Y}^{(2)}}_{\mathrm{chiral,\thinspace apolar}}\nonumber \\
 & + & \underbrace{\mathbf{V}_{i}^{(2s)}\hspace{-0.1cm}\cdot\mathbf{Y}^{(1)}}_{\mathrm{achiral,\,apolar}}\hspace{-0.12cm}-\underbrace{\mathbf{\Omega}_{i}^{\mathcal{A}}\hspace{-0.05cm}\times\hspace{-0.05cm}\bm{\rho}_{i}-\tfrac{1}{60}\boldsymbol{\varepsilon}\cdot\mathbf{V}_{i}^{(4a)}\hspace{-0.12cm}\cdot\mathbf{Y}^{(3)}}_{\mathrm{chiral,\,polar}}\label{eq:slip-truncation}
\end{eqnarray}
have coefficients of polar, apolar and chiral symmetry. Here $\boldsymbol{\varepsilon}$
is the Levi-Civita tensor. The retained modes have physical interpretations:
for a single colloid in an unbounded fluid, $\mathbf{V}^{\mathcal{A}}$
($l\sigma=1s)$ and $\bm{\Omega}^{\mathcal{A}}$ $(l\sigma=2a)$ are
the linear and angular velocities in the absence of external forces
and torques, $\mathbf{V}^{(2s)}$ is the active contribution to the
stresslet, while $\mathbf{V}^{(3a)},\mathbf{V}^{(3t)}$, and $\mathbf{V}^{(4a)}$
are strengths of the chiral torque dipole, polar vector quadrupole,
and chiral octupole respectively. The tensors are parametrized uniaxially,
$\mathbf{V}_{i}^{\mathcal{A}}=v_{s}\mathbf{p}_{i}$, $\mathbf{\boldsymbol{\Omega}}_{i}^{\mathcal{A}}=\omega_{s}\mathbf{p}_{i}$,
$\mbox{\ensuremath{{\bf V}_{i}^{(2s)}=V_{0}^{(2s)}(\mathbf{p}_{i}\mathbf{p}_{i}-\tfrac{\mathbf{I}}{3})}}$
and so on, where $v_{s}$ and $\omega_{s}$ are the speeds of active
translation and rotation and $V_{0}^{(2s)}$ positive (negative) corresponds
to a pusher (puller). \textcolor{black}{The relation of these modes
to exterior fluid flow and Stokes multipoles is explained in }\cite{supplementalM}.

The synthetic active colloids in \cite{palacci2013living} are polar
and achiral (they self-propel but do not spin) while the bacteria
in \cite{petroff2015fast} are polar and chiral (they self-propel
and spin). Both these cases are included in the leading contributions.
In \cite{ghose2014irreducible} a procedure is outlined for estimating
the leading coefficients from experimentally measured flows and it
is shown that the active flow produced by flagellates and green algae
can be modeled by slip. Our model is of sufficient generality, then,
to include both synthetic and biological active colloids, and situations
where swirling and time-dependent slip may be necessary \cite{drescher2010,drescher2011,guasto2010,goldstein2015green}.
\begin{figure*}
\includegraphics[width=0.9\textwidth]{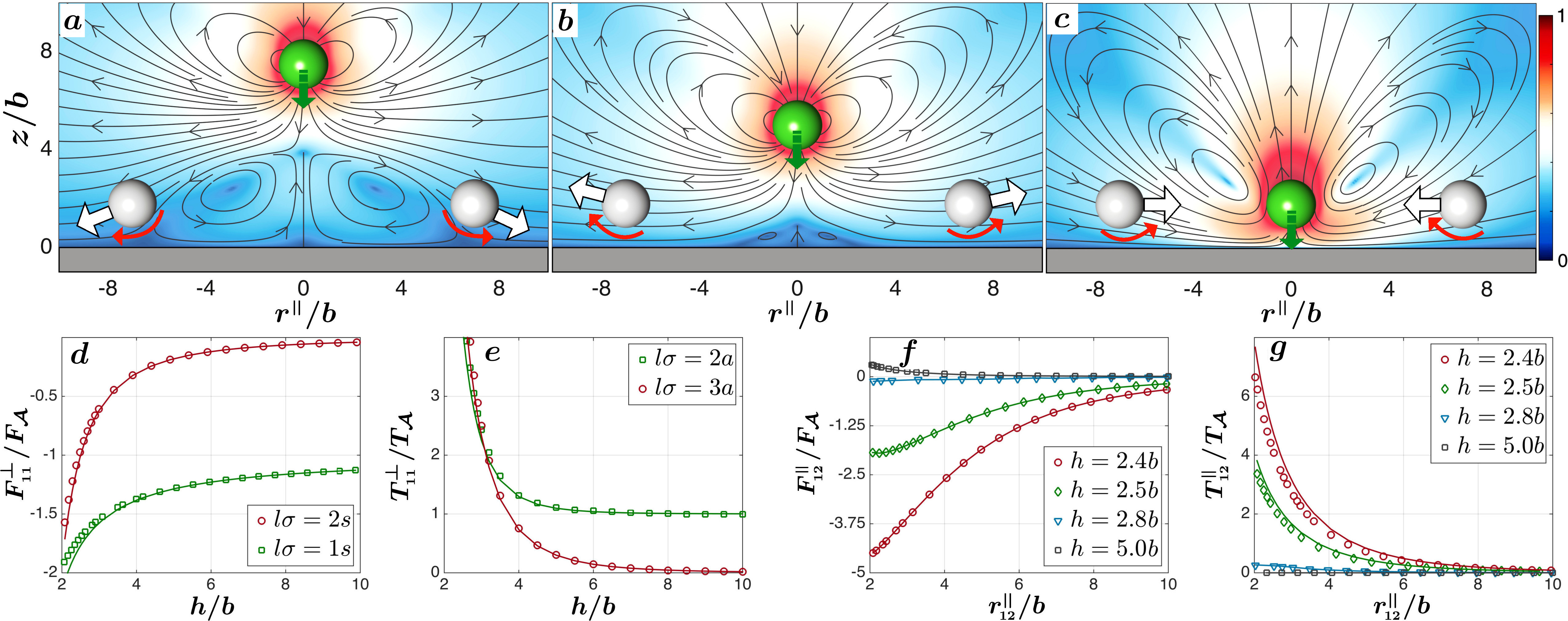} \caption{
    Distortion of the flow produced by leading polar $(l\sigma=1s)$ and
apolar $(l\sigma=2s)$ slip terms in Eq.(\ref{eq:slip-truncation})
as an active colloid of radius $b$, shown in green, approaches a
plane wall. Tracer colloids are show in white. The streamlines of
the fluid flow have been overlaid on the pseudocolor plot of logarithm
of the magnitude of local flow normalised by its maximum. The flow
in (c) results when the colloid is brought to rest near the wall.
Hydrodynamic forces attract nearby colloids, as shown by the thick
white arrows, leading to crystallization. Hydrodynamic torques tend
to reorient the colloids as shown by the curved red arrows. The remaining
graphs show quantitative variation of the active forces and torques
from modes in Eq. (\ref{eq:slip-truncation}) scaled by $F_{\mathcal{A}}=6\pi\eta bv_{s}$
and $T_{\mathcal{A}}=8\pi\eta b^{2}v_{s}$ respectively as a function
of height $h$ of the colloid from the wall and distance, $\mathbf{r}_{ij}=\mathbf{R}_{i}-\mathbf{R}_{j}$,
from other colloids . Solid and dotted lines represent analytical
and numerical results respectively (see text). Here $\parallel$ and
$\perp$ indicate directions parallel and perpendicular to the wall at
$\mathbf{z}=0$.
\label{fig:distortion-of-active-flow}}
\end{figure*}

\emph{Active forces and torques: }Newton's equations of motion for
the colloids reduce, when inertia is negligible, to instantaneous
balance of forces and torques
\begin{equation}
\mathbf{F}_{i}^{H}+\mathbf{F}_{i}^{P}+\boldsymbol{\xi}_{i}^{T}=0,\quad\mathbf{T}_{i}^{H}+\mathbf{T}_{i}^{P}+\boldsymbol{\xi}_{i}^{R}=0.\label{eq:Newton}
\end{equation}
Here $\mathbf{F}_{i}^{H}=\int\mathbf{f}\,d\text{S}_{i}$, $\mathbf{F}^{P}$
and $\mathbf{\boldsymbol{\xi}^{T}}$ are respectively the hydrodynamic,
body and Brownian forces while, $\mathbf{T}_{i}^{H}=\int\boldsymbol{\rho}_{i}\mathbf{\times}\mathbf{f}\,d\text{S}_{i}$,
$\mathbf{T}_{i}^{P}$ and $\boldsymbol{\xi}_{i}^{R}$ are, corresponding
torques, $\boldsymbol{\sigma}$ is the Cauchy stress in the fluid
and $\mathbf{f}=\bm{\hat{\rho}_{i}}\cdot\boldsymbol{\sigma}$ is the
traction. The linearity of the Stokes equation implies that these
must be of the form \begin{subequations}\label{force-formulation}
\begin{alignat}{1}
\mathbf{F}_{i}^{H}= & -\boldsymbol{\gamma}_{ij}^{TT}\mathbf{\cdot V}_{j}-\boldsymbol{\gamma}_{ij}^{TR}\mathbf{\cdot\boldsymbol{\Omega}}_{j}-\sum_{l\sigma=1s}^{\infty}\boldsymbol{\gamma}_{ij}^{(T,\,l\sigma)}\negmedspace\cdot\mathbf{V}_{j}^{(l\sigma)},\label{eq:linear-force-torque}\\
\mathbf{T}_{i}^{H}= & -\boldsymbol{\gamma}_{ij}^{RT}\mathbf{\cdot V}_{j}-\boldsymbol{\gamma}_{ij}^{RR}\mathbf{\cdot\boldsymbol{\Omega}}_{j}-\sum_{l\sigma=1s}^{\infty}\boldsymbol{\gamma}_{ij}^{(R,\,l\sigma)}\negmedspace\cdot\mathbf{V}_{j}^{(l\sigma)},\label{eq:torque-expression}
\end{alignat}
\end{subequations}where repeated particle indices are summed over.
The $\boldsymbol{\gamma}_{ij}^{\alpha\beta}$, with $\alpha,\beta=T,R$,
are the usual friction matrices associated with rigid body motion
and $\mathcal{\boldsymbol{\gamma}}_{ij}^{(\alpha,\,l\sigma)}$ are
friction tensors associated with the irreducible modes of the active
slip. They are of rank $l+1$, $l$, and $l-1$, respectively, for
$\sigma=s,a,t$. The forces and torques depend on relative position
(through the $\mathcal{\boldsymbol{\gamma}}_{ij}^{(\alpha,\,l\sigma)}$)
and on relative orientation (through the $\mathbf{V}_{j}^{(l\sigma)}$).
Their signature under time-reversal shows that the active contributions
are dissipative. 

We calculate the friction tensors using a Galerkin discretization
of the boundary integral equation \cite{singh2014many,singh2016traction}
with the Lorentz-Blake Green's function \cite{blake1971c} which,
by construction, vanishes at the plane wall. The $\boldsymbol{\gamma}_{ij}^{(T,\,l\sigma)}$
decay as $r_{ij}^{-(l+1)}$ and $r_{ij}^{-(l+2)}$ in the directions
parallel and perpendicular to the wall. The $\boldsymbol{\gamma}_{ij}^{(R,\,l\sigma)}$
decay one power of $r_{ij}$ more rapidly. While the force and torque
so obtained are sufficient to study colloidal motion, additional insight
is obtained from studying the flow, which we compute from its boundary
integral representation. Further details are given in \cite{supplementalM}.

The modes $l\sigma=1s$ and $l\sigma=2a$ contribute most dominantly
to forces and torques and they attain their lower bounds far away
from the wall, where their magnitudes are $\textcolor{black}{\ensuremath{F=6\pi\eta bv_{s}}}$
and $T=8\pi\eta b^{3}\omega_{s}$. The bacteria in \cite{petroff2015fast}
have radius $b\sim4\,\mu\text{m}$, swimming speed $v_{s}\sim500\,\mu\text{m/s}$
and angular speed $\omega_{s}\sim50\,s^{-1}$ in a fluid of viscosity
$\eta=10^{-3}\,\text{kg/ms}$. This gives an estimate of $F\sim40\times10^{-12}\,\text{N}$
and$T\sim10^{-16}$ Nm. For the synthetic colloids in \cite{palacci2013living},
$b\sim2\,\mu\text{m}$, $v_{s}\sim10\,\mu\text{m/s}$, which corresponds
to $F_{\mathcal{}}\sim10^{-13}\,\text{N}$. Typical Brownian forces
and torques are of order $\mathcal{O}\left(k_{B}\text{T/b}\right)\sim10^{-15}\,\text{N}$,
and $\mathcal{O}\left(k_{B}\text{T}\right)\sim10^{-21}\,\text{Nm}$
respectively. Thus active forces and torques overwhelm Brownian contributions
by factors of 100 or more in these experiments and, henceforth, we
neglect their effects. Trajectories are obtained by integrating the
kinematic equations $\mathbf{\dot{R}}_{i}=\mathbf{V}_{i}$ and $\dot{\mathbf{p}}_{i}=\mathbf{\Omega}_{i}\times\mathbf{p}_{i}$,
where $\mathbf{V}_{i}$ and $\mathbf{\boldsymbol{\Omega}}_{i}$ satisfy
Eq. (\ref{eq:Newton}) with Brownian contributions removed. Integration
methods and parameter choices are detailed in \cite{supplementalM}.

\emph{Crystallization kinetics: }The kinetics of crystallization obtained
from numerical solutions is shown in Movie 1 \cite{supplementalM},
together with the evolution of the structure factor $S(\mathbf{k})$.
The uniform state is destabilized, most notably for any initial density,
by attractive active hydrodynamic forces. Steric repulsion between
particles balances these to produce crystallites with hexagonal positional
order. Rings in the structure factor first appear at wavenumbers that
correspond to Bragg vectors of the lattice, reminiscent of a spinodal
instability, representing the averaged scattering from randomly oriented
crystallites. These sharpen into Bragg peaks as the crystallites coalesce
and orientational order grows. Finally particles assemble into a single
crystallite which continues to rotate, while the structure factor
shows a clear sixfold symmetry. In Movie 2 \cite{supplementalM} we
show the formation of a hexagonal unit cell from the simulation of
seven polar and chiral active colloids. The crystallite rotates with
an angular velocity parallel to the chiral axis of the colloids. 

\textit{Universal mechanisms:} To better understand the mechanisms
behind active crystallization we show, in Fig. (\ref{fig:distortion-of-active-flow})
, the active flow near a wall and the dominant contributions to the
flow-mediated forces and torques. The top three panels show the increasing
distortion of the flow produced by the leading polar $(l\sigma=1s)$
and apolar $(l\sigma=2s)$ modes for $\mathbf{p}_{i}$ normal to the
wall and $V_{0}^{(2s)}<0.$ The flow develops a monopolar character
as the colloid is brought to rest at a height $h$ by the balance
of hydrodynamic attraction, Fig. (\ref{fig:distortion-of-active-flow}d),
and steric repulsion from the wall. The induced monopole on the colloids
leads to attractive forces between them below a critical height $h$
from the wall as shown in Fig. (\ref{fig:distortion-of-active-flow}f).
Nearby colloids entrained in this flow are attracted towards the central
colloid as shown in the rightmost panel and in Movie 3 \cite{supplementalM}.
The balance of the hydrodynamic attraction and steric repulsion determines
the lattice spacing $d$. We note that even an apolar colloid is attracted
to the wall, Fig. (\ref{fig:distortion-of-active-flow}d), and induces
hydrodynamic attractive forces. Thus, unlike MIPS \cite{cates2015},
polarity is not necessary for crystallization. The induced monopole
also tends to reorient the colloids, by generating a torque in the
plane of wall, as shown by the curved red arrows in Fig. (\ref{fig:distortion-of-active-flow}c)
and quantified in Fig. (\ref{fig:distortion-of-active-flow}g). \textcolor{black}{Their
destabilizing effect can be nullified by external torques $\mathbf{T}_{i}^{P}=T_{0}(\mathbf{\hat{z}}\times\mathbf{p}_{i})$
in the plane of the wall due, for example, to bottom-heaviness. The
orientation can also be stabilized by the chiral terms in Eq. (\ref{eq:slip-truncation}),
which produce torques $\perp$ to the wall, as shown in Fig. (\ref{fig:distortion-of-active-flow}e).
This chiral torque acting $\perp$ to the wall, when combined with
destabilizing torque $\parallel$ to the wall, induces }\textit{\textcolor{black}{active
}}\textcolor{black}{precession of the orientation about the wall normal,
thereby stabilizing the orientations. The role of each of the six
terms in Eq. (\ref{eq:slip-truncation}) in generating positional
order, orientational order and crystal rotation is tabulated in }\textit{\emph{\cite{supplementalM}.
}}Activity \emph{and} body forces pointing away from the wall are
necessary for positional order while bottom-heaviness \emph{or} chirality
is necessary for orientational stability. 

\begin{figure}[t]
\includegraphics[width=0.47\textwidth]{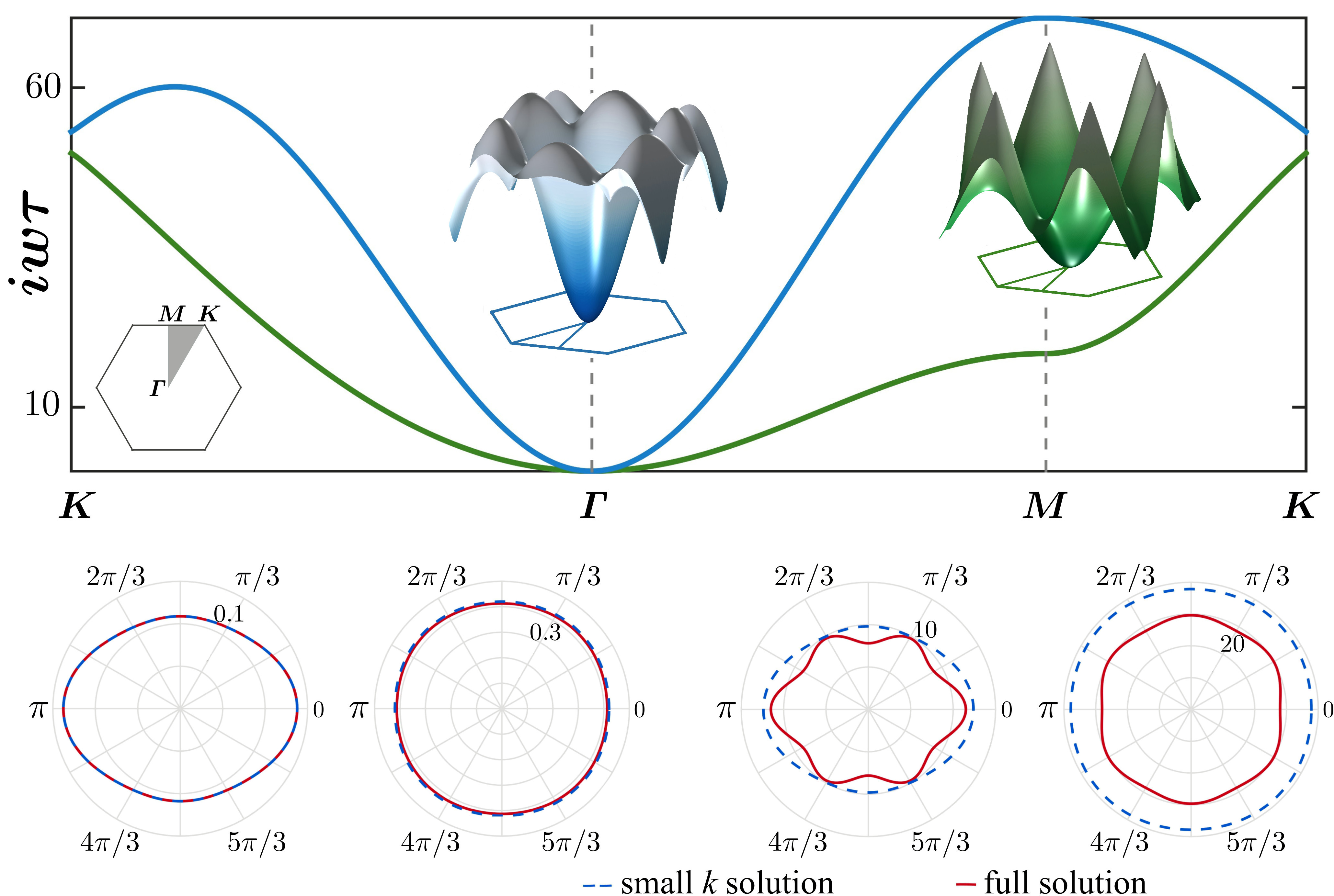}\caption{Branches of the dispersion relation for the two planar normal modes
of relaxation of a hexagonal active crystal. The curves in upper panel
show the dispersion along high symmetry directions of the Brillouin
zone (first inset). The surfaces in the second and third insets show
the dispersion over the entire Brillouin zone. Polar plots in the
lower panel, have comparisons of full numerical solution of Eq. (\ref{eq:det-stability})
with the approximate solution at small $k$ of Eq. (\ref{eq:small-k})
for $k=0.01k_{0}$ (left panel) and $k=0.3k_{0}$ (right panel).\label{fig:Normal-modes}}
\end{figure}

\emph{Harmonic excitations: }We now study harmonic excitations $\mathbf{u}_{i}$
of a perfect hexagonal crystal by expanding the positions as $\mathbf{R}_{i}=\mathbf{R}_{i}^{0}+\mathbf{u}_{i}$
around the stationary state $\mathbf{R}_{i}^{0}=(X_{i}^{0},\,Y_{i}^{0},\,h)$
and ignoring orientational fluctuations. Force balance to leading
order gives 
\begin{eqnarray}
-\boldsymbol{\gamma}_{ij}^{TT}\cdot\dot{\mathbf{u}}_{j} &  & +\left(\mathbf{\boldsymbol{\nabla}}_{j}\boldsymbol{\gamma}_{ij}^{TT}\cdot\mathbf{V}^{\mathcal{A}}-\mathbf{D}_{ij}\right)\cdot\mathbf{u}_{j}=0,\label{eq:stability-eq-ac}
\end{eqnarray}
where $\mathbf{D}_{ij}=-\boldsymbol{\nabla}_{j}\boldsymbol{\nabla}{}_{i}U\big|_{0}$
and $U$ is the sum of all steric potentials.\textit{\emph{ This shows
that relaxation is determined by both activity and elasticity, unlike
in an equilibrium colloidal crystal where elasticity alone relaxes
strains. The normal modes of relaxation can be obtained by Fourier
transforming in the plane and in time. The dispersion is found from
the solutions of }}
\begin{equation}
\det\Big|-i\omega\boldsymbol{\gamma}_{\mathbf{k}}^{TT}+i\mathbf{k}\,\boldsymbol{\gamma}_{\mathbf{k}}^{TT}\cdot\mathbf{V}^{\mathcal{A}}-\mathbf{D}_{\mathbf{k}}\Big|=0.\label{eq:det-stability}
\end{equation}
\textit{\emph{Here $\mathbf{k}=(k_{1},\,k_{2})$ is the wavevector
restricted to the first Brillouin zone \cite{supplementalM}, $\omega$
is the frequency and $\mathbf{D}_{\mathbf{k}}$ }}is the dynamical
matrix\textit{\emph{. The pair of dispersion relations for motion
parallel to the wall are shown in Fig. (\ref{fig:Normal-modes}).
The dispersion for $k\ll k_{0}$, where $k_{0}$ is the magnitude
of the reciprocal lattice vector, is quadratic in wavenumber
\begin{equation}
\omega_{\pm}=-i\tfrac{\gamma_{\perp}^{T}hv_{s}}{2\gamma_{\myparallel}^{T}}f_{\pm}(\theta)\,k^{2},\label{eq:small-k}
\end{equation}
where $f_{\pm}(\theta)$ are angular factors, $\gamma_{\myparallel}^{T}$
and $\gamma_{\perp}^{T}$ are one-body frictions parallel and perpendicular
to the wall, and $\tan\theta=\tfrac{k_{2}}{k_{1}}$. The small-$k$
approximation is compared with the numerical solution in Fig. (\ref{fig:Normal-modes})
and it is found to hold for $k\lesssim0.1k_{0}$. These can be interpreted
as overdamped phonon modes of the active crystal \cite{joannyDiscuss}.
The presence of the active term }}$i\mathbf{k}\,\boldsymbol{\gamma}_{\mathbf{k}}^{TT}\cdot\mathbf{V}^{\mathcal{A}}$
in Eq. (\ref{eq:det-stability}) makes them differ from phonon modes
of a colloidal crystal.

\textit{Discussion}: In this work, we have considered only hydrodynamic
forces and torques, unlike the case of MIPS \cite{tailleur2008statistical,cates2010arrested,cates2013active,cates2015}
where Brownian torques drives reorientations \cite{henkes2011active,fily2012athermal,bialke2012crystallization,redner2013structure}.
We have shown that the latter are at least two orders of magnitude
weaker than the former for experiments in the class of \cite{palacci2013living,petroff2015fast}.
However, it is conceivable that thermal fluctuations will play a more
significant role when the activity is comparatively weak, modifying
both the nature of crystallization transition and the stability of
the crystalline phase. The spinodal-like instability appears due to
the uncompensated long-ranged attractive active forces. These can
be compensated by entropic forces to stabilize the disordered phase
at finite temperatures. A nucleational route to crystallization, with
activity-enhanced rates, is then possible in the regime where the
active forces reduce the nucleation barrier without driving it to
zero. In the crystalline phase, thermal fluctuations will excite both
phonon and topological modes. Phonon fluctuations will destroy long-range
translational order \cite{peierls1935quelques,landau1937theorie},
but due to the activity-enhanced stiffness of these modes, large system
sizes (compared to equilibrium) will be needed to observe the power-law
decay of correlations. Topological defects will be excited at higher
temperatures and a defect unbinding transition \cite{kosterlitz1973ordering,halperin1978theory,nelson1979theory,young1979,chaikin2000principles},
modified by activity, may destroy translational order entirely, producing
instead an \textquotedblleft active\textquotedblright{} hexatic phase.
These present exciting avenues for future research. We remark that
wall-bounded clustering phenomena in algae \cite{drescher2009} and
charged colloids \cite{squires2001effective} are mediated by specific
forms of the universal hydrodynamic mechanisms presented here.

Finally, we suggest that the flow-induced phase separation found here
may provide a paradigm, complementary to MIPS, in which theoretical
and experimental studies of momentum-conserving driven \cite{yeo2015collective}
and active matter \cite{trau1996field,solomentsev1997particle,matas2014hydrodynamic,pandey2014flow,wang2015one,wykes2016dynamic}
may be situated. 

We thank M. E. Cates, P. Chaikin, D. Frenkel, D. J. Pine, A. Laskar
and T. V. Ramakrishnan for helpful discussions and IMSc for computing
resources on the Nandadevi clusters.

\appendix
\vspace{0.6cm}\begin{center}

\textbf{\textcolor{black}{\Large{}Supplemental information}}\end{center}

\section{Active force, torque and flow\label{sec:propulsion-matrices-in-wall-bounded}}

\begin{figure}
\includegraphics[width=0.47\textwidth]{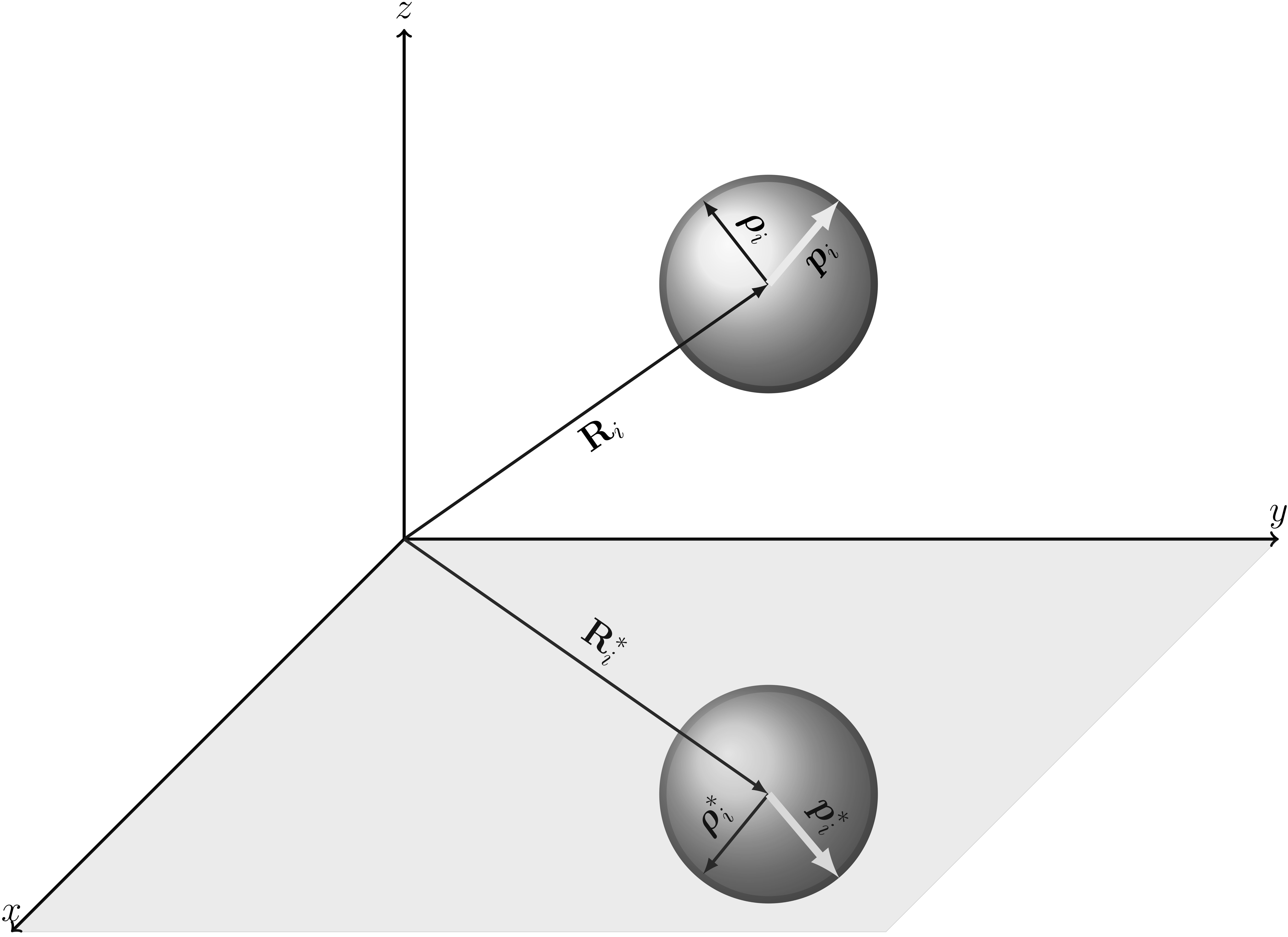}

\caption{Coordinate system used to describe active spherical particles and
its images near a no-slip wall. The $i$-th particle and its image
is shown. See text for description. \label{fig:coordinate-system}}
\end{figure}

We derive, in this section, the expressions for the active forces,
torques, and exterior flow in a suspension of $N$ active colloids
bounded by a plane wall. The system of coordinates is shown in Fig.
(\ref{fig:coordinate-system}). The spheres are centered at $\mathbf{R}_{i}$
and their velocities and angular velocities are $\mathbf{V}_{i}$
and $\mathbf{\Omega}_{i}$ respectively. ${\bf p}_{i}$ denotes the
orientation of the particle while points on the boundaries of the
spheres is given by $\mathbf{s}_{i}=\mathbf{R}_{i}+\boldsymbol{\rho}_{i}$,
where $\boldsymbol{\rho}_{i}$ is the radius vector. To ensure no-slip
on the wall, we associate am image centered at ${\bf R}_{i}^{*}$
with the $i$-th colloid \cite{blake1971c}, and a similar correspondence
for all other quantities of the colloid and its image. 

We closely follows our previous work \cite{singh2014many,singh2016traction}
where a boundary integral formulation has been used to solve the Stokes
equation with arbitrary boundary conditions. The principal difference
here is in the choice of the Green's function which satisfies the
no-slip condition at the plane wall \cite{blake1971c}. In the interest
of being self-contained, we repeat certain key steps en route to the
solution. A clear expression of the linearity of Stokes flow is found
in its integral representation, where the flow in the bulk is given
in terms of integrals of the tractions and velocities at the boundaries
\cite{fkg1930bandwertaufgaben,ladyzhenskaya1969,pozrikidis1992,kim2005},
\begin{alignat}{1}
v_{\alpha}(\mathbf{r})= & -\int G_{\alpha\beta}^{W}(\mathbf{r},\,\mathbf{s}_{j})f_{\beta}(\mathbf{s}_{j})\,d\mathrm{S}_{i}\nonumber \\
+ & \int K_{\beta\alpha\gamma}^{W}(\mathbf{r},\,\mathbf{s}_{j})\hat{\rho}_{\gamma}v_{\beta}(\mathbf{s}_{j})\,d\mathrm{S}_{i},\label{eq:BIE}
\end{alignat}
where repeated particle indices are summed over, $\mathbf{s}_{j}=\mathbf{R}_{j}+\boldsymbol{\rho}_{j}$
is a point on the surface of $j$-th particle and $G_{\alpha\beta}^{W}(\mathbf{r},\,\mathbf{s}_{j})$
is the Green's function of the Stokes system satisfying no-slip condition,
$\mathbf{v}=0$ on the wall at $z=0$. The stress tensor $K_{\alpha\beta\gamma}^{W}(\mathbf{r},\,\mathbf{s}_{j})$
and the pressure vector $P_{\alpha}^{W}$($\mathbf{r},\,\mathbf{s}_{j})$
satisfy $K_{\alpha\beta\gamma}^{W}(\mathbf{r},\,\mathbf{s}_{j})=-\delta_{\alpha\gamma}P_{\beta}^{W}+\eta\left(\nabla_{\gamma}G_{\alpha\beta}^{W}+\nabla_{\alpha}G_{\beta\gamma}^{W}\right)$
and $-\nabla_{\alpha}P_{\beta}^{W}(\mathbf{r},\mathbf{r'})+\eta\nabla^{2}G_{\alpha\beta}^{W}=-\delta\left(\mathbf{r}-\mathbf{r'}\right)\delta_{ij}$
respectively. 

We solve the Fredholm integral equation of Eq. (\ref{eq:BIE}) by
expanding the boundary fields in irreducible tensorial spherical harmonics,
$\mathbf{Y}^{(l)}$, which are orthogonal basis function on the surface
of the sphere $\frac{1}{4\pi b^{2}}\int\mathbf{Y}^{(l)}(\widehat{\bm{\rho}})\,\mathbf{Y}^{(l')}(\widehat{\bm{\rho}})\,d\mathrm{S}=\delta_{ll'}\,\frac{l!\,(2l-1)!!}{(2l+1)}\mathbf{\Delta}^{(l)},$
where $\mathbf{\Delta}^{(l)}$ is tensor of rank $2l$, projecting
any $l$-th order tensor to its symmetric irreducible form \cite{mazur1982,hess1980formeln}.
The boundary velocity including active slip and its expansion in this
basis has been provided above. The orthogonality of the basis functions
gives the expansion coefficients in terms of surface integrals of
traction and velocity as \cite{ladd1988,ghose2014irreducible},
\begin{alignat}{1}
\mathbf{F}_{i}^{(l)} & =\frac{1}{(l-1)!(2l-3)!!}\int\mathbf{f}(\mathbf{R}_{i}+\bm{\rho}_{i})\mathbf{Y}^{(l-1)}(\bm{\hat{\rho}}_{i})\,d\mathrm{S}_{i},\nonumber \\
\mathbf{V}_{i}^{(l)} & =\frac{2l-1}{4\pi b^{2}}\int\mathbf{v}^{\mathcal{A}}(\mathbf{R}_{i}+\bm{\rho}_{i})\mathbf{Y}^{(l-1)}(\bm{\hat{\rho}}_{i})\,d\mathrm{S}_{i}.
\end{alignat}
The coefficients of the traction and velocity are tensors of rank
$l$ and can be written as irreducible tensor of rank $l,$ $l-1$
and $l-2$ \cite{singh2014many}. The first term in the traction expansion
is the force $\mathbf{F}_{i}^{(1)}=\mathbf{F}_{i}^{H}$, while the
antisymmetric part of the second term is the torque $b\boldsymbol{\varepsilon}\cdot\mathbf{F}_{i}^{(2)}=\mathbf{T}_{i}^{H}$.
The first term in the velocity expansion is $\mathbf{V}_{i}^{(1)}=-\mathbf{V}_{i}^{\mathcal{A}}$
and the antisymmetric part of the second term is $\tfrac{1}{2b}\boldsymbol{\boldsymbol{\varepsilon}}\cdot\mathbf{V}_{i}^{(2)}=-\mathbf{\Omega}_{i}^{\mathcal{A}}$.
Here $\mathbf{V}_{i}^{\mathcal{A}}=-\tfrac{1}{4\pi b^{2}}\int\mathbf{v}^{\mathcal{A}}(\bm{\rho}_{i})\,d\mathrm{S}_{i}$
denotes the self-propulsion, while $\bm{\Omega}_{i}^{\mathcal{A}}=-\frac{3}{8\pi b^{4}}\int\bm{\rho}_{i}\times\mathbf{v}^{\mathcal{A}}(\bm{\rho}_{i})\,d\mathrm{S}_{i}$
denotes the self-rotation of an isolated active colloid in unbounded
flow. The expression for fluid flow can be obtained in terms of coefficients
of traction and velocity, 
\begin{alignat}{1}
\mathbf{v}(\mathbf{r})=\sum_{l=1}^{\infty}\Big(-\boldsymbol{G}_{j}^{(l)}\cdot\mathbf{F}_{j}^{(l)}+ & \boldsymbol{K}_{j}^{(l)}\cdot\mathbf{V}_{j}^{(l)}\Big),\label{eq:fluid-flow}
\end{alignat}
where the boundary integrals $\boldsymbol{G}_{j}^{(l)}$ and $\boldsymbol{K}_{j}^{(l)}$
can be written in terms of Green's function and its derivatives (Appendix
\ref{sec:Expression-for-boundary}). We multiply the fluid velocity
by the $l$-th tensorial harmonic and integrate over the $i$-th boundary.
Using the orthogonality of these basis functions, we obtain an infinite-dimensional
linear system of equations for the unknown traction coefficients \cite{singh2014many},
\begin{alignat}{1}
\tfrac{1}{2}\,\mathbf{V}_{i}^{(l)}=\sum_{l'=1}^{\infty}\Big(-\boldsymbol{G}_{ij}^{(l,\,l')}\cdot\mathbf{F}_{j}^{(l')} & +\boldsymbol{K}_{ij}^{(l,\,l')}\cdot\mathbf{V}_{j}^{(l')}\Big),\label{eq:linear-system}
\end{alignat}
where the matrix elements $\boldsymbol{G}_{ij}^{(l,\,l')}$ and $\boldsymbol{K}_{ij}^{(l,\,l')}$
can be evaluated in terms of the Green's function and its derivatives,
as given in Appendix \ref{sec:Expression-for-boundary}. 

The traction and velocity coefficients are reducible and their irreducible
decomposition is given as \cite{brunn1976effect,schmitz1980force},
\begin{eqnarray*}
\mathbf{F}_{i}^{(ls)}=\overbracket[0.7pt][2.0pt]{\mathbf{F}_{i}^{(l)}},\quad & \mathbf{F}_{i}^{(la)}=\overbracket[0.7pt][2.0pt]{\bm{\varepsilon}\cdot\mathbf{F}_{i}^{(l)}},\quad & \mathbf{F}_{i}^{(lt)}={\bm{\delta}\cdot\mathbf{F}_{i}^{(l)}},\\
\mathbf{V}_{i}^{(ls)}=\overbracket[0.7pt][2.0pt]{\mathbf{V}_{i}^{(l)}},\quad & \mathbf{V}_{i}^{(la)}=\overbracket[0.7pt][2.0pt]{\boldsymbol{\mathbf{\varepsilon}}\cdot\mathbf{V}_{i}^{(l)}},\quad & \mathbf{V}_{i}^{(lt)}={\bm{\delta}\cdot\mathbf{V}_{i}^{(l)}}.
\end{eqnarray*}
Here the operator $\overbracket[0.7pt][2.0pt]{(\dots)}=\boldsymbol{\Delta}^{(l)}(\dots)$
extracts the symmetric irreducible part of the tensor it acts on.
We use these irreducible coefficients and the linear system of equations
to solve for the unknown traction in terms of the known boundary velocity
\cite{singh2016traction}. The relations between the irreducible coefficients
of the traction and velocity, then, becomes \cite{singh2016traction},
\begin{alignat}{1}
\mathbf{F}_{i}^{(l\sigma)}= & -\boldsymbol{\gamma}_{ij}^{(l\sigma,\,1s)}\cdot\mathbf{V}_{j}-\boldsymbol{\gamma}_{ij}^{(l\sigma,\,R)}\cdot\mathbf{\Omega}_{j}\nonumber \\
- & \sum_{l'\sigma'=1s}^{\infty}\boldsymbol{\gamma}_{ij}^{(l\sigma,\,l'\sigma')}\cdot\mathbf{V}_{j}^{(l'\sigma')}.\label{eq:main-traction-l}
\end{alignat}
This infinite set of equations, called the traction laws \cite{singh2016traction},
manifestly shows the linear relation between the traction and velocity
coefficients and defines the friction tensors. The expressions for
the friction tensors can be obtained by an iterative scheme \cite{singh2016traction}.
We use the one-body solution as the initial guess for the iteration,
\begin{equation}
\mathbf{F}_{i}^{H}=-\gamma^{T}(\mathbf{V}_{i}-\mathbf{V}_{i}^{\mathcal{A}}),\qquad\mathbf{T}_{i}^{H}=-\gamma^{R}(\mathbf{\Omega}_{i}-\mathbf{\Omega}_{i}^{\mathcal{A}}),
\end{equation}
where $\gamma^{T}$ and $\gamma^{R}$ are one particle friction corresponding
to translation and rotation. Near a wall no-slip wall, they are, ${\gamma}^{T}=\gamma_{\perp}^{T}\hat{\boldsymbol{z}}+\gamma_{\myparallel}^{T}(\hat{\boldsymbol{x}}+\hat{\boldsymbol{y}})$
and ${\gamma}^{R}=\gamma_{\perp}^{R}\hat{\boldsymbol{z}}+\gamma_{\myparallel}^{R}(\hat{\boldsymbol{x}}+\hat{\boldsymbol{y}})$
\cite{kim2005}. Here $\myparallel$ and $\perp$ subscripts indicating
directions parallel and perpendicular to the wall. The expressions
after one iteration, corresponding to the ``first reflection'' in
Smoluchowski's classical method, are shown in Appendix \ref{appendix:Evaluation-of-gamma}.

\section{Crystalline steady states \label{sec:Crystalline-steady-states}}

In this section we work out the steady states of the active crystals
using the leading terms of the force and torque equations. Using the
leading order force balance for $i$-th particle, the steady state
condition for position is given as 
\begin{eqnarray}
\bm{\gamma}_{ij}^{TT}\cdot\mathbf{V}_{j}^{\mathcal{A}}+\mathbf{F}_{i}^{P}=0.\label{eq:minimal-steady-state}
\end{eqnarray}
Here $\mathbf{V}_{i}^{\mathcal{A}}=-v_{s}\hat{\mathbf{z}}$ is self-propulsion
of the colloid at a speed $v_{s}$, assumed to be moving $\perp$
to the wall. The body force $\mathbf{F}_{i}^{P}=-\boldsymbol{\nabla}_{{\scriptscriptstyle \mathbf{R}_{i}}}U$
is due to a short-ranged repulsive potential $U,$ which depends on
displacement $\mathbf{r}_{ij}=\mathbf{R}_{i}-\mathbf{R}_{j}$ and
is given as, $U(r_{ij})=\epsilon\left(\frac{r_{min}}{r_{ij}}\right)^{12}-2\epsilon\left(\frac{r_{min}}{r_{ij}}\right)^{6}+\epsilon,$
for $r_{ij}<r_{min}$ and zero otherwise \cite{weeks1971role}, where
$\epsilon$ is the potential strength. The same potential has been
used to model colloid-colloid repulsion $\mathbf{F}_{i}^{PP}$ and
the colloid-wall repulsive force $\mathbf{F}_{i}^{PW}$.

\begin{figure}[t]
\includegraphics[width=0.47\textwidth]{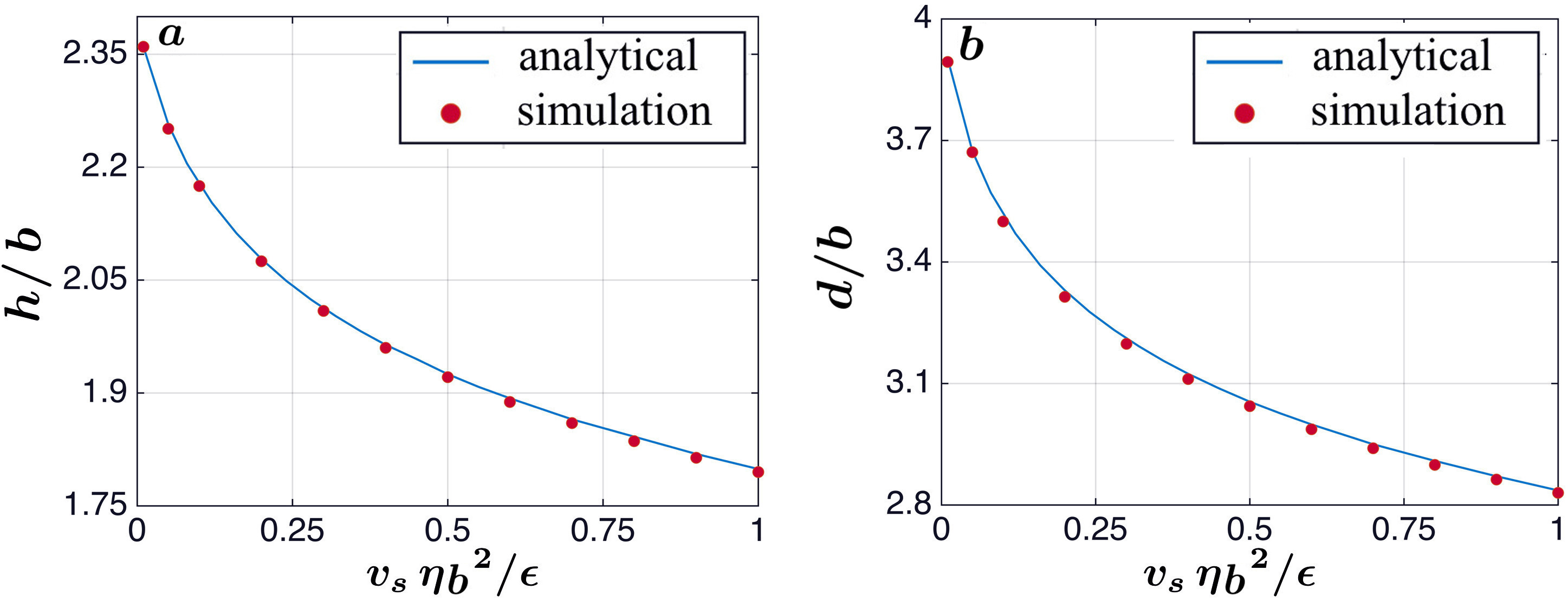}\caption{Steady states of active crystallization. Left panel has the plot of
leading terms for the analytical solution of height $h$, shown in
solid line, along with the full numerical result, shown as dotted
curve. Right panel has similar set of plots for lattice spacing $d$.
The leading order estimates are found to be in agreement with the
numerical solution.\label{fig:steady-state}}
\end{figure}

\textit{One- and two-body dynamics:} To estimate the height at which
the particle is brought to rest close to the wall, we use the $z-$component
of the force balance, $-\gamma_{\perp}^{T}v_{s}=F_{3}^{PW}$. Here
$F_{3}^{PW}$ is the repulsive force from the wall in $\mathbf{\hat{z}}$
direction, while $\gamma_{\perp}^{T}v_{s}$ is the attractive force
of the colloid to the wall in same direction. The balance between
the attraction and repulsion sets the height $h$ at which the colloid
is brought to rest. We now consider force balance for a pair of particles
in planar direction,
\begin{alignat}{1}
-v_{s}\gamma_{\myparallel}^{T}\gamma_{\perp}^{T}\mbox{\ensuremath{\mathcal{F}}}_{i}^{0}\mbox{\ensuremath{\mathcal{F}}}_{j}^{0}G_{\alpha3}^{W}(\mathbf{R}_{i},\mathbf{R}_{j})+F_{\alpha}^{PP} & =0,
\end{alignat}
where 
\begin{equation}
\mathcal{F}_{i}^{l}=\left(1+\tfrac{b^{2}}{4l+6}\boldsymbol{\nabla}_{{\scriptscriptstyle \mathbf{R}_{i}}}^{2}\right),
\end{equation}
 is an operator encoding the finite size of the sphere and $\alpha$
may takes either of the values 1 or 2 corresponding to two equivalent
directions parallel to wall. We have used results provided in Appendix
\ref{appendix:Evaluation-of-gamma}, to write the expression for friction.
The solution of this equation gives the lattice spacing $d$. For
fixed particle-wall potential, increasing $v_{s}$ decreases the resting
height $h$ and separation between\emph{ pairs, $d$,} as show in
Fig. (\ref{fig:steady-state}). 

\textit{Rotational dynamics:} In Fig. (\ref{fig:Dynamics-of-crystallization}),
we show the state diagram, obtained from simulation, which shows that
the crystal is stable over a critical strengths of either bottom-heaviness
or chirality. For an initially symmetric distribution, a crystal stabilized
by external torque alone \emph{does not rotate}, while the crystal
\emph{stabilized by chirality does rotate}. When the crystal is rotating
at an angular velocity $\mathbf{\Omega}_{c}$ about its center of
mass $\mathbf{R}_{c}$, the velocity the $i$-th colloid at position
$\mathbf{R}_{i}$ can be then written as $\mathbf{\dot{R}}_{i}=\mathbf{\Omega}_{c}\times\mathbf{R}_{i}$.
Force balance parallel to the wall is then
\begin{eqnarray}
\bm{\gamma}_{ij}^{TT}\cdot\left[\mathbf{\Omega}_{c}\times(\mathbf{R}_{j}-\mathbf{R}_{0}^{c})\right]+\bm{\gamma}_{ij}^{TR}\cdot\mathbf{\Omega}_{j}=0.\label{eq:minimal-steady-state-1}
\end{eqnarray}
The angular speed perpendicular to wall is $\Omega=\Omega_{i}^{\mathcal{A}}$.
This implies that in absence of chiral self-rotation there is \textit{no}
rotation of the crystal. The angular velocity of the crystal can be
obtained by power counting - $\bm{\gamma}_{ij}^{TT}$ scales as $r_{ij}^{-3}$
in direction parallel to wall while $\bm{\gamma}_{ij}^{TR}$ scales
as $r_{ij}^{-4}$. The angular velocity of the crystal, then, scales
as $\Omega_{c}\propto1/R_{c}^{2}$. In Fig. (\ref{fig:Dynamics-of-crystallization})
we show that rotation period of a crystal scales inversely as number
of particles $N$ in the crystal for an assembly of chiral particles,
which is an excellent agreement with a recent experiment \cite{petroff2015fast}.

\section{Harmonic excitations\label{sec:dispersion-relation} }

In this section we study harmonic excitations $\mathbf{u}_{i}$ of
the crystal about a stationary state $\mathbf{R}_{i}^{0}=(X_{i}^{0},\,Y_{i}^{0},\,h)$,
such that $\dot{\mathbf{R}}_{i}^{0}=0$ and $\mathbf{\Omega}_{i}^{0}=0$
at this location. The force balance condition is then $\boldsymbol{\gamma}_{ij}^{TT}\cdot\mathbf{V}_{j}^{\mathcal{A}}\big|_{0}+\mathbf{F}_{i}^{P}\big|_{0}=0$.
We, now, consider a small displacement about this state $\mathbf{R}_{i}=\mathbf{R}_{i}^{0}+\mathbf{u}_{i}$.
Expanding the friction tensors about the stationarity point, we have,
\[
\boldsymbol{\gamma}_{ij}^{TT}=\boldsymbol{\gamma}_{ij}^{TT}\big|_{0}+(\mathbf{u}_{i}\cdot\mathbf{\boldsymbol{\nabla}}_{{\scriptscriptstyle \mathbf{R}_{i}}}-\mathbf{u}_{j}\cdot\mathbf{\boldsymbol{\nabla}}_{{\scriptscriptstyle \mathbf{R}_{j}}})\mathbf{}\boldsymbol{\gamma}_{ij}^{TT}\big|_{0}+\mathcal{O}(\mathbf{u}^{2}).
\]
The force can be expanded in a similar way $\mathbf{F}_{i}^{P}=\mathbf{F}_{i}^{P}\big|_{0}-\mathbf{D}_{ij}\cdot\mathbf{u}_{j},$
where $\mathbf{D}_{ij}=-\boldsymbol{\nabla}_{{\scriptscriptstyle \mathbf{R}_{j}}}\mathbf{\boldsymbol{\nabla}}_{{\scriptscriptstyle \mathbf{R}_{i}}}U\big|_{0}$.
Using the equations of motion and considering terms linear in the
displacement, the equation becomes 
\begin{eqnarray*}
-\boldsymbol{\gamma}_{ij}^{TT}\cdot\dot{\mathbf{u}}_{j} &  & +\left(\mathbf{\boldsymbol{\nabla}}_{{\scriptscriptstyle \mathbf{R}_{j}}}\boldsymbol{\gamma}_{ij}^{TT}\cdot\mathbf{V}^{\mathcal{A}}-\mathbf{D}_{ij}\right)\cdot\mathbf{u}_{j}=0.
\end{eqnarray*}
We seek a solution of the form $\mathbf{u}_{i}(t)=\mathbf{u}_{\mathbf{k}}(t)\,e^{i\mathbf{k}\cdot\mathbf{R}_{i}}.$
Using this, the force balance condition becomes,
\begin{equation}
-\boldsymbol{\gamma}_{\mathbf{k}}^{TT}\cdot\dot{\mathbf{u}}_{k}+\left(i\mathbf{k}\,\boldsymbol{\gamma}_{\mathbf{k}}^{TT}\cdot\mathbf{V}^{\mathcal{A}}-\mathbf{D}_{\mathbf{k}}\right)\cdot\mathbf{u}_{\mathbf{k}}=0.\label{eq:stability-fourier}
\end{equation}
Here $\mathbf{D}_{\mathbf{k}}$ is the Fourier transform of $\mathbf{D}_{ij}$
and $\boldsymbol{\gamma}_{\mathbf{k}}^{TT}$ is the Fourier transform
of the friction tensor
\begin{flalign*}
\mathbf{D}_{\mathbf{k}} & =\sum_{i=1}^{N}\mathbf{D}_{i1}\,e^{i\mathbf{k}\cdot(\mathbf{R}_{i}-\mathbf{R}_{1})},\\
\boldsymbol{\gamma}_{\mathbf{k}}^{TT} & =\sum_{i=1}^{N}\boldsymbol{\gamma}_{i1}^{TT}\,e^{i\mathbf{k}\cdot(\mathbf{R}_{i}-\mathbf{R}_{1})}.
\end{flalign*}
Here, $\mathbf{D}_{\mathbf{k}}$ is called the dynamical matrix \cite{born1954dynamical}.
We now write $\boldsymbol{\gamma}_{i1}^{TT}$ in terms of its planar
Fourier transform 
\[
\boldsymbol{\gamma}_{i1}^{TT}=\int\hat{\boldsymbol{\gamma}}_{\mathbf{k}}^{TT}(\mathbf{k};\,h)e^{-i\mathbf{k}'\cdot(\mathbf{R}_{i}-\mathbf{R}_{1})}\,\frac{d^{2}k'}{(2\pi)^{2}},
\]
to obtain an expression for $\boldsymbol{\gamma}_{\mathbf{k}}^{TT}$,
\begin{eqnarray}
\boldsymbol{\gamma}_{\mathbf{k}}^{TT}(\mathbf{k};\,h) & = & \sum_{i}\int\hat{\boldsymbol{\gamma}}_{\mathbf{k}}^{TT}(\mathbf{k};\,h)e^{-i(\mathbf{k}'-k)\cdot(\mathbf{R}_{i}-\mathbf{R}_{1})}\,\frac{d^{2}k'}{(2\pi)^{2}},\nonumber \\
 & = & \frac{1}{A_{c}}\sum_{\lambda}\hat{\boldsymbol{\gamma}}_{\mathbf{k}}^{TT}(\mathbf{k}+\mathbf{q}_{\lambda};\,h).
\end{eqnarray}
Here we have used the identity 
\begin{equation}
\sum_{i}e^{-i\mathbf{k}\cdot\mathbf{R}_{i}}=\frac{(2\pi)^{2}}{A_{c}}\sum\mathbf{\boldsymbol{\delta}}(\mathbf{k}-\mathbf{q}_{\lambda}),
\end{equation}
where $A_{c}$ is area of the unit cell and $\mathbf{q}_{\lambda}$
are reciprocal lattice vectors. We now identify two parts of $\boldsymbol{\gamma}_{\mathbf{k}}^{TT},$
\begin{alignat*}{1}
 & \boldsymbol{\gamma}_{\mathbf{k}}^{TT}(\mathbf{k};\,h)=\hat{\boldsymbol{\gamma}}_{{\scriptscriptstyle \mathbf{k}_{0}}}^{TT}(\mathbf{k};\,h)+\sum_{\lambda'}\hat{\boldsymbol{\gamma}}_{{\scriptscriptstyle \mathbf{k}_{q}}}^{TT}(\mathbf{k}+\mathbf{q}_{\lambda};\,h)
\end{alignat*}
Here $\hat{\boldsymbol{\gamma}}_{{\scriptscriptstyle \mathbf{k}_{0}}}^{TT}(\mathbf{k};\,h)$
corresponds to the $\mathbf{q}_{\lambda}=0$ and terms at arbitrary
non-zero $q$ are denoted by $\hat{\boldsymbol{\gamma}}_{{\scriptscriptstyle \mathbf{k}_{q}}}^{TT}(\mathbf{k};\,h)$.
Their leading order forms can be written as
\begin{alignat*}{1}
\hat{\boldsymbol{\gamma}}_{{\scriptscriptstyle \mathbf{k}_{0}}}^{TT}(\mathbf{k};\,h) & =\gamma^{T}\mathbf{I}+\frac{\gamma^{T}\gamma^{T}}{A_{c}}\mathcal{F}^{k}\,\hat{\mathbf{G}}^{W}(\mathbf{k};\,h),\\
\hat{\boldsymbol{\gamma}}_{{\scriptscriptstyle \mathbf{k}_{q}}}^{TT}(\mathbf{k};\,h) & =\frac{\gamma^{T}\gamma^{T}}{A_{c}}\mathcal{F}^{k}\,\sum_{\lambda'}\hat{\mathbf{G}}^{W}(\mathbf{k}+\mathbf{q}_{\lambda};\,h).
\end{alignat*}
Here $\mathcal{F}^{k}=1-b^{2}k^{2}/3$ and $\hat{\mathbf{G}}^{W}(\mathbf{k};\,h)$
is the two-dimensional Fourier transform of $\mathbf{G}^{W}$ (see
Appendix \ref{appendix:FT-GW}). The prime on the summation on the
right indicates that $\lambda=0$ is excluded from the sum. We now
turn to the calculation of the dynamical matrix, 
\begin{flalign*}
\mathbf{D}_{\mathbf{k}} & =\sum_{i=1}^{N}\left(\frac{\mathbf{I}}{r^{2}}U'+\frac{\mathbf{r}\,\mathbf{r}}{r^{4}}U''\right)_{0}\left(1-e^{i\mathbf{k}\cdot\mathbf{R}_{i}}\right).
\end{flalign*}
Here $U'=-12\epsilon\,\big[\left(r_{min}/d\right)^{12}-\left(r_{min}/d\right)^{6}\big]$
and $U''=12\epsilon\,\big[14\left(r_{min}/d\right)^{12}-8\left(r_{min}/d\right)^{6}\big]$.
We evaluate the above in the nearest neighbor approximation in the
direction parallel to the wall. The expression for $\mathbf{D}_{\mathbf{k}}$
and $\boldsymbol{\gamma}_{\mathbf{k}}^{TT}$ can be evaluated numerically
by summing over the reciprocal lattice vectors. The sum is unconditionally
and rapidly convergent as the Green's function decays as $r_{ij}^{-3}$
in the direction parallel to the wall. The dispersion is obtained
numerically from Eq. (\ref{eq:stability-fourier}) and is shown in
Fig. (\ref{fig:Normal-modes}).

\textit{Long-wavelength approximation: }Analytical expression for
the normal modes can be obtained in the $k\rightarrow0$ limit. Keeping
terms of the $\mathcal{O}(k^{2})$, Eq. (\ref{eq:stability-fourier})
becomes
\begin{eqnarray*}
\,\left(\begin{array}{c}
\dot{u}_{k_{1}}\\
\dot{u}_{k_{2}}
\end{array}\right) & = & -\frac{h\gamma_{\perp}^{T}v_{s}\,k^{2}}{\gamma_{\myparallel}^{T}}\boldsymbol{f}(\theta)\left(\begin{array}{c}
u_{k_{1}}\\
u_{k_{2}}
\end{array}\right),
\end{eqnarray*}
\[
\boldsymbol{f}(\theta)=\left(\begin{array}{cc}
c_{1}\cos^{2}\theta+\mathcal{C}_{2}\sin^{2}\theta & c_{3}\sin\theta\cos\theta\\
c_{3}\sin\theta\cos\theta & c_{4}\sin^{2}\theta+c_{5}\cos^{2}\theta
\end{array}\right).
\]
Here $k_{1}=k\cos\theta$, $k_{2}=k\sin\theta$ and $c_{i}$ are positive
constants that can be determined in terms of the parameters of the
steric potential and the friction tensors:	 $c_{1}=\gamma_{\perp}^{T}h/2\eta A_{c}+\left(\tfrac{3}{2}U'+\tfrac{9}{8}U''+\right)/h\gamma_{\myparallel}^{T}v_{s},$
$c_{2}=\left(\tfrac{3}{2}U'+\tfrac{3}{8}U''+\right)/h\gamma_{\myparallel}^{T}v_{s},\,$
$c_{3}=\gamma_{\perp}^{T}h/2\eta A_{c}+\tfrac{3}{4}U''/h\gamma_{\myparallel}^{T}v_{s}$,
$c_{4}=\gamma_{\perp}^{T}h/2\eta A_{c}+3c_{5}$ and $c_{5}=\left(\tfrac{1}{2}U'+\tfrac{3}{8}U''+\right)/h\gamma_{\myparallel}^{T}v_{s}$.
We can now diagonalize this matrix equation to obtain the relaxation
of the overdamped modes after Fourier transforming in time. The eigenvalues
of the resulting equations give the dispersion relation \textit{\emph{
\begin{equation}
\omega_{\pm}=-i\tfrac{\gamma_{\perp}^{T}hv_{s}}{2\gamma_{\myparallel}^{T}}f_{\pm}(\theta)\,k^{2},\label{eq:small-k-1}
\end{equation}
in terms of an angular factor,}} $f_{\pm}(\theta)=c_{15}\cos^{2}\theta+c_{24}\sin^{2}\theta\pm[(c_{15}\cos^{2}\theta+c_{24}\sin^{2}\theta)^{2}-4(c_{1}c_{5}\cos^{4}\theta+c_{2}c_{4}\sin^{4}\theta-c_{3}^{2}\cos^{2}\theta\sin^{2}\theta+(c_{1}c_{4}+c_{2}c_{5})\cos^{2}\theta\sin^{2}\theta)]{}^{1/2}$,
with $c_{15}=c_{1}+c_{5}$ and $c_{24}=c_{2}+c_{4}$. The comparison
of the long wavelength solution with the full numerical solution has
been plotted in Fig. (\ref{fig:Normal-modes}). For \textit{\emph{$k\lesssim0.1k_{0}$,}}
the approximate solution shows excellent agreement.

\section{Numerical method}

In this section, we outline the method used to simulate the dynamics
of active colloidal particles near a no-slip wall. We invert Eq. (\ref{eq:main-traction-l})
to obtain rigid body motions in terms of the known slip modes, body
forces and torques \cite{singh2016traction}. This gives the ``mobility''
formulation,
\begin{alignat*}{1}
\mathsf{\mathbf{V}}_{i} & =\boldsymbol{\mu}_{ij}^{TT}\cdot\mathbf{F}_{j}^{P}+\boldsymbol{\mu}_{ij}^{TR}\cdot\mathbf{T}_{j}^{P}+\sum_{l\sigma=2s}^{\infty}\boldsymbol{\pi}_{ij}^{(T,l\sigma)}\cdot\mathbf{\mathsf{\mathbf{V}}}_{j}^{(l\sigma)}+\mathsf{\mathbf{V}}_{i}^{\mathcal{A}},\\
\mathsf{\mathbf{\Omega}}_{i} & =\boldsymbol{\mu}_{ij}^{RT}\cdot\mathbf{F}_{j}^{P}+\boldsymbol{\mu}_{ij}^{RR}\cdot\mathbf{T}_{j}^{P}+\sum_{l\sigma=2s}^{\infty}\boldsymbol{\pi}_{ij}^{(R,\,l\sigma)}\cdot\mathbf{\mathsf{\mathbf{V}}}_{j}^{(l\sigma)}+\mathsf{\mathbf{\Omega}}_{i}^{\mathcal{A}}.
\end{alignat*}
The mobility matrices $\boldsymbol{\mu}_{ij}^{\alpha\beta}$, with
($\alpha,\beta=T,R$), are inverses of the friction matrices $\boldsymbol{\gamma}_{ij}^{\alpha\beta}$
\cite{kim2005}. The propulsion tensors $\boldsymbol{\pi}_{ij}^{(\alpha,\,l\sigma)}$,
first introduced in \cite{singh2014many}, relate the rigid body motion
to modes of the active velocity. They are related to the slip friction
tensors by \cite{singh2016traction},
\begin{alignat*}{1}
-\bm{\pi}_{ij}^{(\text{T},\,l\sigma)} & =\boldsymbol{\mu}_{ik}^{TT}\cdot\boldsymbol{\gamma}_{kj}^{(T,\,l\sigma)}+\boldsymbol{\mu}_{ik}^{TR}\cdot\boldsymbol{\gamma}_{kj}^{(R,\,l\sigma)},\\
-\bm{\pi}_{ij}^{(R,\,l\sigma)} & =\boldsymbol{\mu}_{ik}^{RT}\cdot\boldsymbol{\gamma}_{kj}^{(T,\,l\sigma)}+\boldsymbol{\mu}_{ik}^{RR}\cdot\boldsymbol{\gamma}_{kj}^{(R,\,l\sigma)}.
\end{alignat*}
We retain modes corresponding to $l\sigma=1s,\,2s,2a,\,3a,\,3t$ and
$4a$ in the active slip. The role of these individual modes is summarized
in Table (\ref{tab:slip-mode-verbose}). The mobilities are calculated
using the PyStokes \cite{pystokes} library. The initial distribution
of particles is chosen to be the random packing of hard-spheres \cite{skoge2006packing}.
We use an adaptive time step integrator using the backward differentiation
formula (BDF) to integrate these equations of motion \cite{langtangen2012tutorial}.
In Table (\ref{tab:Table-of-simulation-parameters}) of Appendix,
we present the parameters used to generate the figures. 

\begin{figure*}
\selectlanguage{american}%
\centering\includegraphics[width=0.82\textwidth]{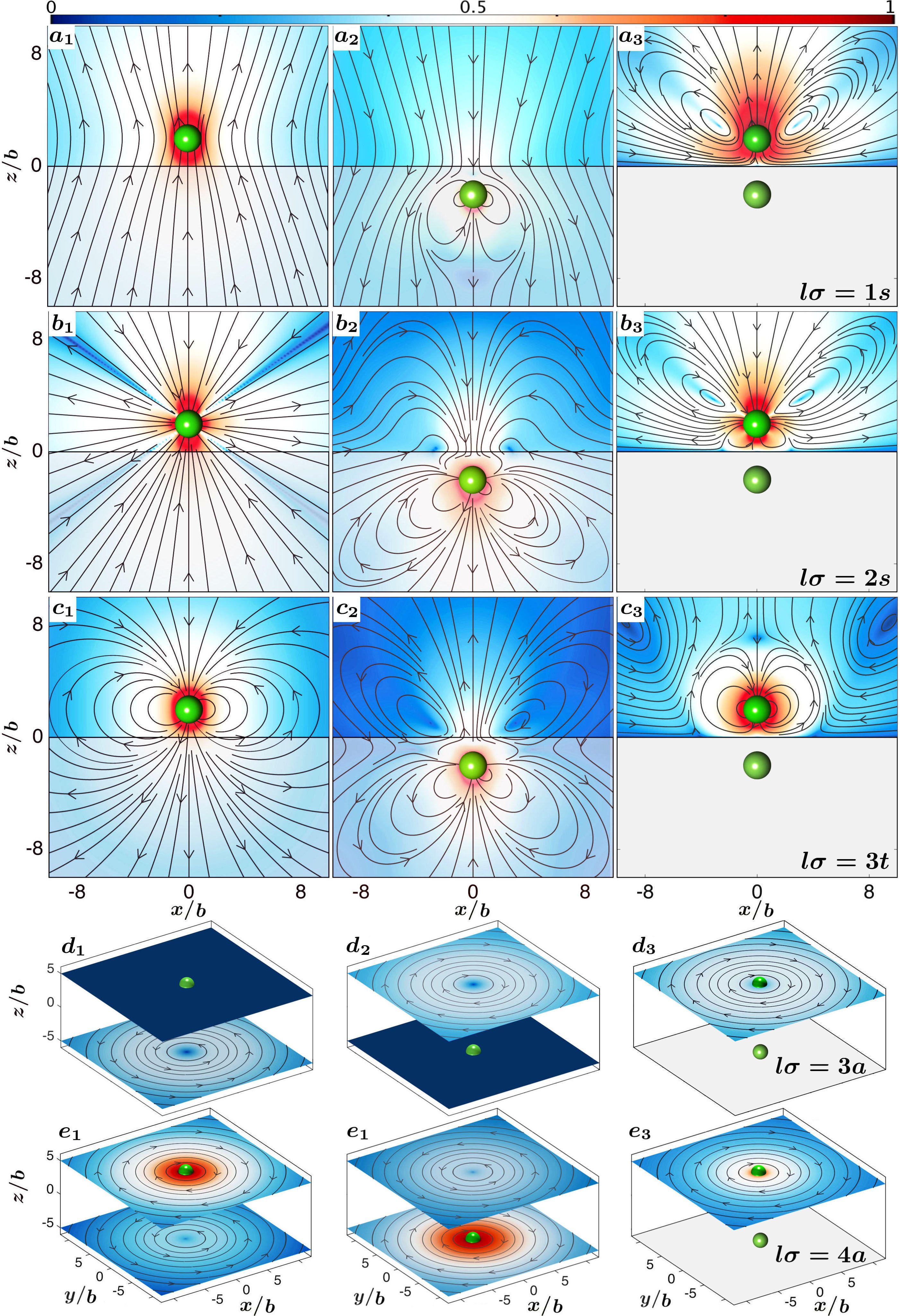}

\selectlanguage{english}%
\caption{Irreducible axisymmetric and swirling components of fluid flow induced
by an active colloid at a height $h$ from the wall. The streamlines
of the fluid flow have been overlaid on the pseudo-color plot of the
normalized logarithmic flow speed. The first column is the flow due
to source alone while second column has the flow from the image, and
their sum is plotted in the third column for all the irreducible modes,
panel (a)-(e), used in this work. The first two rows show flow produced
by a force monopole and a force dipole respectively. Third row is
the flow due to a vector quadrupole, while the last two rows (panel
d and e) are the swirling flows due to a torque dipole and antisymmetric
octupole respectively. The torque-dipole and octupole induces a net
rotation of colloids near a wall. The orientation of the colloid,
in all these plots, is chosen to be along the wall normal. A linear
combination of panel (a-c) has been used to plot the Fig. 1(a)-(c)
.\label{fig:flowDistortion-near-a-wall}}
\end{figure*}

\textcolor{black}{As the $\mathbf{V}_{i}^{(l\sigma)}$ are irreducible
tensors, it is natural to parametrize them in terms of the tensorial
spherical harmonics. The uniaxial parametrizations used here are:
$\mathbf{V}_{i}^{(2s)}=V_{0}^{(2s)}\mathbf{Y}^{(2)}(\mathbf{p})$,
$\mathbf{V}_{i}^{(3t)}=V_{0}^{(3t)}\mathbf{Y}^{(1)}(\mathbf{p})$,
$\mathbf{V}_{i}^{(3a)}=V_{0}^{(3a)}\mathbf{Y}^{(2)}(\mathbf{p})$
and $\mathbf{V}_{i}^{(4a)}=V_{0}^{(4a)}\mathbf{Y}^{(3)}(\mathbf{p})$
where
\begin{align*}
Y_{\alpha}^{(1)}( & \mathbf{p})=p_{\alpha},\qquad Y_{\alpha\beta}^{(2)}(\mathbf{p})=p_{\alpha}p_{\beta}-\tfrac{1}{3}\delta_{\alpha\beta},\\
Y_{\alpha\beta\gamma}^{(3)}(\mathbf{p}) & =p_{\alpha}p_{\beta}p_{\gamma}-\tfrac{1}{5}[p_{\alpha}\delta_{\beta\gamma}+p_{\beta}\delta_{\alpha\gamma}+p_{\gamma}\delta_{\alpha\beta}].
\end{align*}
The flow due to the modes retained in the minimal truncation are shown
in Fig. (\ref{fig:flowDistortion-near-a-wall}). The force-free motion
of a sphere far away from wall is described by the mode $\mathbf{V}_{i}^{\mathcal{A}}$
or $\mathbf{V}_{i}^{(3t)}$. Though near a wall more interesting things
are expected, which is especially relevant for our study, as near
a wall, the image flow from modes other than $\mathbf{V}_{i}^{\mathcal{A}}$
or $\mathbf{V}_{i}^{(3t)}$ can lead to particle motion. In particular,
the $\mathbf{V}_{i}^{(2s)}$ mode, whose far field is that of a symmetric
irreducible dipole, can produce motion near a wall, due to image interactions.
This is clear from the streamlines in panel }\textit{\textcolor{black}{\emph{(b)}}}\textcolor{black}{\emph{
}}\textcolor{black}{of Fig. (\ref{fig:flowDistortion-near-a-wall}).}
\begin{table}[H]
\centering\renewcommand{\arraystretch}{2} 

\begin{tabular}{|>{\centering}b{1.4cm}|>{\centering}b{2cm}|>{\centering}b{2cm}|>{\centering}b{2cm}|}
\hline 
Slip mode & Positional clustering & Orientational stability & Cluster rotation \tabularnewline
\hline 
\hline 
$\mathbf{V}^{a}$ & Yes & No & No\tabularnewline
\hline 
$\mathbf{\Omega}^{a}$ & No & Yes & No\tabularnewline
\hline 
$\mathbf{V}^{(2s)}$ & Yes & No & No\tabularnewline
\hline 
$\mathbf{V}^{(3t)}$ & Yes & No & No\tabularnewline
\hline 
$\mathbf{V}^{(3a)}$ & No & Yes & Yes\tabularnewline
\hline 
$\mathbf{V}^{(4a)}$ & No & Yes & Yes\tabularnewline
\hline 
\end{tabular}\caption{\label{tab:slip-mode-verbose}\textcolor{black}{Role of different
terms in truncation of the slip expansion as given in Eq. (\ref{eq:slip-truncation}).
It should be noted the rotation column is for a completely symmetric
cluster. A cluster with any degree of asymmetry will rotate, irrespective
of stability by bottom-heaviness or chirality. }}
\end{table}
\widetext 
\begin{table*}
\renewcommand{\arraystretch}{2} 

\begin{tabular}{|>{\centering}b{2cm}|c|>{\centering}p{2cm}|>{\centering}p{2cm}|>{\centering}p{2cm}|c|c|}
\hline 
Figure & \# of colloids & $v_{s}$ & $V_{0}^{(3a)}/v_{s}$ & $T_{0}/\epsilon$ & Wall WCA & Inter-particle WCA\tabularnewline
\hline 
\hline 
1 (1-f) & 1024 & 0.1 & 100 & 0 & $r_{min}=3.4b$,$\,$$\epsilon=0.083$ & $r_{min}=5b$,$\,$$\epsilon_{p}=0.004$\tabularnewline
\hline 
1 (g) & - & 0.01 & 100 & 0 & $r_{min}=3.4b$,$\,$$\epsilon=0.083$ & $r_{min}=5b$,$\,$$\epsilon_{p}=0.004$\tabularnewline
\hline 
1 (h) & 256 & 0.01 & - & - & $r_{min}=3.4b$,$\,$$\epsilon=0.083$ & $r_{min}=5b$,$\,$$\epsilon_{p}=0.004$\tabularnewline
\hline 
2  & 1 and 2 & 0.1 & 100 & - & - & -\tabularnewline
\hline 
3 & - & 0.1 & - & - & - & $r_{min}=5b$,$\,$$\epsilon_{p}=0.004$\tabularnewline
\hline 
\end{tabular}\caption{\label{tab:Table-of-simulation-parameters}Simulation parameters used
to study the active crystallization. Throughout the paper: radius
of particle $b$ =1, $\eta=1/6$, the strength of the modes, $V_{0}^{(2s)}=0.01$
and $V_{0}^{(3t)}=1$. $\mathsf{\mathbf{\Omega}}_{i}^{\mathcal{A}}$
and $V_{0}^{(4a)}$ are non-zero only for Fig. (2e), where they are
of unit strength. All the simulations are in three space dimensions. }
\end{table*}

\section{Expression for boundary integrals and matrix elements\label{sec:Expression-for-boundary}}

The boundary integrals in fluid flow, Eq. (\ref{eq:fluid-flow}),
can be solved exactly. The resulting solution is given in terms of
the Green's function, Eq. (\ref{eq:wall-G}), and its derivatives
\cite{singh2014many},
\begin{alignat*}{1}
\boldsymbol{G}_{j}^{(l)}(\mathbf{r},\mathbf{R}_{j}) & =\frac{2l-1}{4\pi b^{2}}\int\mathbf{G}^{W}(\mathbf{r},\mathbf{R}_{j}+\bm{\rho}_{j})\mathbf{Y}^{(l-1)}(\hat{\bm{\rho}}_{j})\,d\mathrm{S}_{i}=b^{l}\mathcal{F}^{l-1}\mathbf{\bm{\nabla}}_{{\scriptscriptstyle \mathbf{R}_{j}}}^{(l-1)}\mathbf{G}^{W}(\mathbf{r},\mathbf{R}_{j}),\\
\boldsymbol{K}_{j}^{(l)}(\mathbf{r},\mathbf{R}_{j}) & =\frac{1}{(l-1)!(2l-3)!!}\int\mathbf{K}^{W}(\mathbf{r},\mathbf{R}_{j}+\bm{\rho}_{j})\cdot\mathbf{n}\mathbf{Y}^{(l-1)}(\hat{\bm{\rho}}_{j})\,d\mathrm{S}_{i}=\frac{4\pi b^{l+1}}{(l-2)!(2l-1)!!}\mathcal{F}^{l-1}\mathbf{\bm{\nabla}}_{{\scriptscriptstyle \mathbf{R}_{j}}}^{(l-2)}\mathbf{K}^{W}(\mathbf{r},\mathbf{R}_{j}).
\end{alignat*}
The integrals appearing in the linear system of the equations, Eq.
(\ref{eq:linear-system}), are,
\begin{alignat*}{1}
\boldsymbol{G}_{ij}^{(l,\,l')}(\mathbf{R}_{i},\mathbf{R}_{j}) & =\frac{(2l-1)(2l'-1)}{(4\pi b^{2})^{2}}\int\mathbf{Y}^{(l-1)}(\hat{\bm{\rho}}_{i})\mathbf{G}^{W}(\mathbf{R}_{i}+\bm{\rho}_{i},\,\mathbf{R}_{j}+\bm{\rho}_{j})\mathbf{Y}^{(l'-1)}(\hat{\bm{\rho}}_{j})\,d\mathrm{S}_{i}\,d\mathrm{S}_{j}\\
\boldsymbol{K}_{ij}^{(l,\,l')}(\mathbf{R}_{i},\mathbf{R}_{j}) & =\frac{2l-1}{4\pi b^{2}\,(l-1)!(2l-3)!!}\int\mathbf{Y}^{(l-1)}(\hat{\bm{\rho}}_{i})\mathbf{K}^{W}(\mathbf{R}_{i}+\bm{\rho}_{i},\,\mathbf{R}_{j}+\bm{\rho}_{j})\cdot\mathbf{n}\mathbf{Y}^{(l'-1)}(\hat{\bm{\rho}}_{j})\,d\mathrm{S}_{i}\,d\mathrm{S}_{j}.
\end{alignat*}
These integrals are solved exactly to give matrix elements in terms
of the Green's function and its derivatives \cite{singh2014many},
\begin{alignat*}{1}
\boldsymbol{G}_{ij}^{(l,\,l')}(\mathbf{R}_{i},\mathbf{R}_{j}) & =\begin{cases}
{\displaystyle \mathcal{G}_{ii}^{(l,\,l')}+b^{l+l'-2}\mathcal{F}_{i}^{l-1}\mathcal{F}_{j}^{l'-1}\bm{\nabla}_{{\scriptscriptstyle \mathbf{R}_{i}}}^{(l-1)}\bm{\nabla}_{{\scriptscriptstyle \mathbf{R}_{j}}}^{(l'-1)}\mathbf{G}^{*}(\mathbf{R}_{i},\mathbf{R}_{j});} & \qquad\qquad\qquad\qquad{\displaystyle \qquad\quad}{\displaystyle j=i,}\\
{\displaystyle b^{l+l'-2}\mathcal{F}_{i}^{l-1}\mathcal{F}_{j}^{l'-1}\bm{\nabla}_{{\scriptscriptstyle \mathbf{R}_{i}}}^{(l-1)}\bm{\nabla}_{{\scriptscriptstyle \mathbf{R}_{j}}}^{(l'-1)}\mathbf{G}^{W}(\mathbf{R}_{i},\mathbf{R}_{j});} & \qquad\qquad\qquad\qquad{\displaystyle \qquad\quad{\displaystyle j\neq i},}
\end{cases}\\
\boldsymbol{K}_{ij}^{(l,\,l')}(\mathbf{R}_{i},\mathbf{R}_{j}) & =\begin{cases}
-{\displaystyle \tfrac{1}{2}\delta_{ll'}\bm{\Delta}^{(l-1)}+\frac{4\pi b^{(l+l'-1)}}{(l'-2)!(2l'-1)!!}\mathcal{F}_{i}^{l-1}\mathcal{F}_{j}^{l'-1}\bm{\nabla}_{{\scriptscriptstyle \mathbf{R}_{i}}}^{(l-1)}\bm{\nabla}_{{\scriptscriptstyle \mathbf{R}_{j}}}^{(l'-2)}\mathbf{K}^{*}(\mathbf{R}_{i},\mathbf{R}_{j});}\quad\qquad\qquad & {\displaystyle j=i,}\\
{\displaystyle \frac{4\pi b^{(l+l'-1)}}{(l'-2)!(2l'-1)!!}\mathcal{F}_{i}^{l-1}\mathcal{F}_{j}^{l'-1}\bm{\nabla}_{{\scriptscriptstyle \mathbf{R}_{i}}}^{(l-1)}\bm{\nabla}_{{\scriptscriptstyle \mathbf{R}_{j}}}^{(l'-2)}\mathbf{K}^{W}(\mathbf{R}_{i},\mathbf{R}_{j});}\qquad\qquad\qquad\qquad\qquad & {\displaystyle j\neq i,}
\end{cases}
\end{alignat*}
\[
\mathcal{G}_{ii}^{(l,\,l')}=\delta_{ll'}\frac{2l-1}{2\pi b}\int\mathbf{Y}^{(l-1)}(\hat{\bm{\rho}})\left(\mathbf{\bm{I}}-\hat{\bm{\rho}}\hat{\bm{\rho}}\right)\mathbf{Y}^{(l-1)}(\hat{\bm{\rho}})\,d\Omega.
\]

\section{First order off-diagonal approximation for friction tensors\label{appendix:Evaluation-of-gamma}}

The expressions for the friction tensor can be calculated from the
solution of the linear system, provided above, using the Jacobi method
\cite{singh2016traction}. The first order approximation to friction
tensors used in this work are provide below,
\begin{eqnarray*}
\Big(\boldsymbol{\gamma}_{ij}^{(TT)} & \Big)^{[1]}= & \gamma^{T}\gamma^{T}\,\mathcal{F}_{i}^{0}\mathcal{F}_{j}^{0}\,\mathbf{G}^{W}(\mathbf{R}_{i},\mathbf{R}_{j}),\quad\Big(\boldsymbol{\gamma}_{ij}^{(RT)}\Big)^{[1]}=\tfrac{1}{2}\gamma^{T}\gamma^{R}\,\boldsymbol{\nabla}_{{\scriptscriptstyle \mathbf{R}_{i}}}\times\mathbf{G}^{W}(\mathbf{R}_{i},\mathbf{R}_{j}),
\end{eqnarray*}
\[
\Big(\boldsymbol{\gamma}_{ij}^{(TR)}\Big)^{[1]}=\tfrac{1}{2}\gamma^{T}\gamma^{R}\,\boldsymbol{\nabla}_{{\scriptscriptstyle \mathbf{R}_{j}}}\times\left(\mathbf{G}^{W}(\mathbf{R}_{i},\mathbf{R}_{j})\right),\quad\Big(\boldsymbol{\gamma}_{ij}^{(RR)}\Big)^{[1]}=\tfrac{1}{4}\gamma^{R}\gamma^{R}\,\boldsymbol{\nabla}_{{\scriptscriptstyle \mathbf{R}_{i}}}\times\left(\boldsymbol{\nabla}_{{\scriptscriptstyle \mathbf{R}_{j}}}\times\mathbf{G}^{W}(\mathbf{R}_{i},\mathbf{R}_{j})\right),
\]
\[
\Big(\boldsymbol{\gamma}_{ij}^{(T,\,2s)}\Big)^{[1]}=\frac{28\pi\eta b^{2}}{3}\gamma^{T}\,\mathcal{F}_{i}^{0}\mathcal{F}_{j}^{1}\,\boldsymbol{\nabla}_{{\scriptscriptstyle \mathbf{R}_{j}}}\mathbf{G}^{W}(\mathbf{R}_{i},\mathbf{R}_{j}),\quad\Big(\boldsymbol{\gamma}_{ij}^{(R,\,2s)}\Big)^{[1]}=\frac{28\pi\eta b^{2}}{6}\gamma^{R}\,\boldsymbol{\nabla}_{{\scriptscriptstyle \mathbf{R}_{i}}}\times\left(\boldsymbol{\nabla}_{{\scriptscriptstyle \mathbf{R}_{j}}}\mathbf{G}^{W}(\mathbf{R}_{i},\mathbf{R}_{j})\right),
\]
\[
\Big(\boldsymbol{\gamma}_{ij}^{(T,\,3a)}\Big)^{[1]}=\frac{13\pi\eta b^{3}}{9}\gamma^{T}\,\boldsymbol{\nabla}_{{\scriptscriptstyle \mathbf{R}_{j}}}(\boldsymbol{\nabla}_{{\scriptscriptstyle \mathbf{R}_{j}}}\times\mathbf{G}^{W}(\mathbf{R}_{i},\mathbf{R}_{j})),\quad\Big(\boldsymbol{\gamma}_{ij}^{(R,\,3a)}\Big)^{[1]}=\frac{13\pi\eta b^{3}}{18}\gamma^{R}\boldsymbol{\nabla}_{{\scriptscriptstyle \mathbf{R}_{i}}}\times\left(\boldsymbol{\nabla}_{{\scriptscriptstyle \mathbf{R}_{j}}}(\boldsymbol{\nabla}_{{\scriptscriptstyle \mathbf{R}_{j}}}\times\mathbf{G}^{W}(\mathbf{R}_{i},\mathbf{R}_{j}))\right),
\]
\[
\Big(\boldsymbol{\gamma}_{ij}^{(T,\,3t)}\Big)^{[1]}=-\frac{4\pi\eta b^{3}}{5}\gamma^{T}\,\boldsymbol{\nabla}_{{\scriptscriptstyle \mathbf{R}_{j}}}^{2}\mathbf{G}^{W}(\mathbf{R}_{i},\mathbf{R}_{j}),\quad\Big(\boldsymbol{\gamma}_{ij}^{(T,\,4a)}\Big)^{[1]}=-\frac{121\pi\eta b^{4}}{10}\gamma^{T}\boldsymbol{\nabla}_{{\scriptscriptstyle \mathbf{R}_{j}}}\boldsymbol{\nabla}_{{\scriptscriptstyle \mathbf{R}_{j}}}(\boldsymbol{\nabla}_{{\scriptscriptstyle \mathbf{R}_{j}}}\times\mathbf{G}^{W}(\mathbf{R}_{i},\mathbf{R}_{j})),
\]
\[
\Big(\boldsymbol{\gamma}_{ij}^{(R,\,4a)}\Big)^{[1]}=-\frac{121\pi\eta b^{4}}{20}\gamma^{R}\boldsymbol{\nabla}_{{\scriptscriptstyle \mathbf{R}_{i}}}\times\left(\boldsymbol{\nabla}_{{\scriptscriptstyle \mathbf{R}_{j}}}\boldsymbol{\nabla}_{{\scriptscriptstyle \mathbf{R}_{j}}}(\boldsymbol{\nabla}_{{\scriptscriptstyle \mathbf{R}_{j}}}\times\mathbf{G}^{W}(\mathbf{R}_{i},\mathbf{R}_{j}))\right),\quad\Big(\boldsymbol{\gamma}_{ij}^{(R,\,3t)}\Big)^{[1]}=0.
\]

\section{Fourier transform of the Lorentz-Blake Green's function\label{appendix:FT-GW}}

In this section, we derive the Fourier transform of the Green's function
for a fluid flow bounded by a plane infinite wall. Blake \cite{blake1971c}
has derived the Green function of the Stokes equation which satisfies
no-slip condition on the wall, 
\begin{alignat}{1}
G_{\alpha\beta}^{\text{w}}(\mathbf{R}_{i},\,\mathbf{R}_{j}) & =G_{\alpha\beta}(\mathbf{R}_{i},\,\mathbf{R}_{j})+G_{\alpha\beta}^{*}(\mathbf{R}_{i},\,\mathbf{R}_{j}^{*})=G_{\alpha\beta}(\mathbf{R}_{i},\,\mathbf{R}_{j})-G_{\alpha\beta}(\mathbf{R}_{i},\,\mathbf{R}_{j})+G'_{\alpha\beta}(\mathbf{R}_{i},\,\mathbf{R}_{j})\label{eq:wall-G}\\
 & =G_{\alpha\beta}(\mathbf{r}_{ij}^{*})-G_{\alpha\beta}(\mathbf{r}_{ij}^{*})-2h\nabla_{{\scriptscriptstyle \mathbf{r}_{ij}^{*}}}G_{\alpha3}(\mathbf{r}_{ij}^{*})\mathcal{M}_{\beta\gamma}+h^{2}\nabla_{{\scriptscriptstyle \mathbf{r}_{ij}^{*}}}^{2}G_{\alpha\gamma}(\mathbf{r}_{ij}^{*})\mathcal{M}_{\beta\gamma}.\nonumber 
\end{alignat}
Here $\mathbf{r}_{ij}=\mathbf{\mathbf{R}}_{i}-\mathbf{\mathbf{R}}_{j}$,
$\mathbf{r}_{ij}^{*}=\mathbf{\mathbf{R}}_{i}-\mathbf{\mathbf{R}}_{j}^{*}$
and $\boldsymbol{\mathcal{M}}=\mathbf{I}-2\mathbf{\hat{z}}\mathbf{\hat{z}}$.
$\mathbf{G}$ is the Green's function in the unbounded fluid flow,
\[
\mathbf{G}(\mathbf{R}_{i},\,\mathbf{R}_{j})=\frac{1}{8\pi\eta}\left(\frac{\mathbf{I}}{r_{ij}}+\frac{\mathbf{r}_{ij}\mathbf{r}_{ij}}{r_{ij}^{3}}\right).
\]
We define the Fourier transform in the plane of the wall as, 
\[
\hat{\varphi}(k_{1},\,k_{2},\,r_{3})=\mathbb{\mathbb{F}}\left[\varphi\right]=\frac{1}{(2\pi)^{2}}\int\varphi(r_{1},\,r_{2},\,r_{3})e^{i(k_{1}r_{1}+k_{2}r_{2})}\,dr_{1}dr_{2}.
\]
 The Fourier transform of $G'_{ij}$, last term of Eq. (\ref{eq:wall-G}),
is then \cite{blake1971c}, 
\[
\hat{G}'_{\alpha\beta}(\mathbf{k};\,h)=\frac{h}{2\eta k}\left[ik_{\alpha_{1}}(\delta_{\alpha3}\delta_{j\alpha_{1}}+\delta_{\beta3}\delta_{\alpha\alpha_{1}})+h\left(ikk_{\alpha_{1}}\{\delta_{\alpha3}\delta_{j\alpha_{1}}-\delta_{\beta3}\delta_{\alpha\alpha_{1}}\}-k_{\alpha_{1}}k_{\alpha_{2}}\delta_{\alpha\alpha_{1}}\delta_{\beta\alpha_{2}}-\delta_{\alpha3}\delta_{\beta3}k^{2}\right)\right]e^{-2kh},
\]
where $\alpha_{1}$ and $\alpha_{2}$ only take values 1 or 2 corresponding
to directions parallel to wall. The rest of terms in Eq. (\ref{eq:wall-G}),
can be transformed using the relation $\mathbb{\mathbb{F}}\left[\frac{1}{r}\right]=\frac{2\pi\,e^{-kz}}{k}$.
The two-dimensional Fourier transform of the wall Green's function
for a source at height $h$ from the wall is then, with $\mathcal{E}=1-e^{-2kh}$,
\begin{eqnarray}
\hat{\mathbf{G}}^{W}(\mathbf{k};\,h) & =\frac{1}{4\eta k^{3}}\left(\begin{array}{ccc}
\mathcal{E}k_{2}^{2}+2hkk_{1}^{2}e^{-2kh}\quad & -\mathcal{E}k_{1}k_{2}-2hkk_{1}k_{2}e^{-2kh}\quad & -i2hk^{2}k_{1}e^{-2kh}\\
\\
-\mathcal{E}k_{1}k_{2}-2hkk_{1}k_{2}e^{-2kh}\quad & \mathcal{E}k_{1}^{2}+2hkk_{2}^{2}e^{-2kh}\quad & -i2hk^{2}k_{2}e^{-2kh}\\
\\
-i2hk^{2}k_{1}e^{-2kh}\quad & -i2hk^{2}k_{1}e^{-2kh}\quad & \mathcal{E}k^{2}-2hk^{3}e^{-2kh}
\end{array}\right)+\hat{\mathbf{G}}'(\mathbf{k};\,h).\label{eq:Gw-FT}
\end{eqnarray}

\begin{thebibliography}{54}
\bibitem [{sli()}]{slipConstraint}%
  \BibitemOpen
  \href@noop {} { {\bibinfo {title} {The active slip may have arbitrary
  radial and tangential components, but the surface can not be a source or
  sink}}\ }\BibitemShut {NoStop}%
\end{thebibliography}
\end{document}